\documentclass[reqno,12pt]{article}
\usepackage[english]{babel}

\usepackage{cite}

\usepackage{ae} 
\usepackage[T1]{fontenc}
\usepackage{amsmath}
\usepackage{amssymb}
\usepackage{mathrsfs}
\usepackage{graphicx}
\usepackage{mdframed}
\usepackage{color}
\definecolor{darkblue}{cmyk}{0.9,0.9,0,0}
\usepackage[colorlinks=true,linkcolor=darkblue,citecolor=darkblue,urlcolor=darkblue]{hyperref}
\usepackage{epsfig}

\usepackage{graphicx}
\usepackage{tikz}

\newcommand{\beq}{\begin{equation}}
\newcommand{\eeq}{\end{equation}}
\newcommand\beqa{\begin{eqnarray}}
\newcommand\eeqa{\end{eqnarray}}
\newcommand\bea{\begin{array}}
\newcommand\eea{\end{array}}

\def\XXint#1#2#3{{\setbox0=\hbox{$#1{#2#3}{\int}$}
\vcenter{\hbox{$#2#3$}}\kern-.5\wd0}}

\newcommand{\nn}{\nonumber}

\newcommand{\neqa}{\nonumber\end{eqnarray}}
\newcommand{\la}[1]{\label{#1}}

\renewcommand{\d}{\partial}

\newcommand{\<}{{\langle}}
\renewcommand{\>}{{\rangle}}

\newcommand{\re}{\relax{\rm I\kern-.18em R}}

\renewcommand{\sp}{p\hspace{-.40em}/}

\def\su2{{SU(2)}}

\def\a{{\alpha}}

\def\[{\left[}
\def\]{\right]}

\def\a{\alpha}

\def\({\left(}
\def\){\right)}
\def\[{\left[}
\def\]{\right]}

\def\<{\langle}
\def\>{\rangle}

\def\i2{\frac{i}{2}}

\def\spi{\relax{\rm \pi\kern-0.5em /}}
\def\sA{\relax{\rm A\kern-0.5em /}}
\def\sp{\relax{\rm p\kern-0.5em /}}
\def\sd{\relax{\rm \d\kern-0.5em /}}
\def\sk{\relax{\rm k\kern-0.5em /}}
\def\sn{\relax{\rm n\kern-0.5em /}}
\def\sl{\relax{\rm l\kern-0.5em /}}
\def\sP{\relax{\rm P\kern-0.7em /}}
\def\sBethe{\relax{\rm \Bethe\kern-0.5em /}}

\newcommand{\ie}{{\it i.e.,\ }}
\newcommand{\eg}{{\it e.g.,\ }}
\newcommand{\mt}[1]{\textrm{\tiny #1}}
\newcommand{\GN}{G_\mt{N}}
\newcommand{\reef}[1]{(\ref{#1})}
\newcommand{\bz}{{\bar z}}
\newcommand{\bZ}{{\bar Z}}

	\def\i{\textrm{i}}

\usepackage{varioref}
\usepackage{makeidx}
\makeindex

\newcommand\blfootnote[1]{%
  \begingroup
  \renewcommand\thefootnote{}\footnote{\hspace{-6mm}#1}%
  \addtocounter{footnote}{-1}%
  \endgroup
}

\numberwithin{equation}{section}

\begin{document}

\thispagestyle{empty}

\renewcommand{\thefootnote}{\fnsymbol{footnote}}
\setcounter{page}{1}
\setcounter{footnote}{0}
\setcounter{figure}{0}

\vspace{-0.4in}

\begin{center}
$$$$
{\Large
Correlation Functions of Huge Operators in AdS$_3$/CFT$_2$:\\ 
Domes, Doors and Book Pages
\par}
\vspace{1.0cm}

{Jacob Abajian,$^\text{\tiny 1,\tiny 2}$ Francesco Aprile,$^\text{\tiny 3}$ Robert C. Myers,$^\text{\tiny 1}$ Pedro Vieira$^\text{\tiny 1,\tiny 4}$}
\blfootnote{
\tt{jabajian@pitp.ca, faprile@ucm.es, pedrogvieira@gmail.com, rmyers@pitp.ca}}
\\ \vspace{1.2cm}
\footnotesize{
{\it $^\text{\tiny 1}$Perimeter Institute for Theoretical Physics,
Waterloo, Ontario N2L 2Y5, Canada  }\\
{\it $^\text{\tiny 2}$Department of Physics \& Astronomy, University of Waterloo, Waterloo, Ontario N2L 3G1, Canada }\\ 
{\it $^\text{\tiny 3}$Dept.~de F\'isica Te\'orica \& IPARCOS, Ciencias F\'isicas,
Universidad Complutense, 28040 Madrid, Spain}\\
{\it $^\text{\tiny 4}$ICTP South American Institute for Fundamental Research, IFT-UNESP, S\~ao Paulo, SP Brazil 01440-070}
\vspace{4mm}
}
\end{center}


\begin{abstract}

We describe solutions of asymptotically AdS$_3$ Einstein gravity  
that are sourced by the insertion of operators in the 
boundary CFT$_2$, whose dimension scales with the central charge of the theory. 
%
%
Previously, we found that the geometry corresponding to a black hole two-point function 
is simply related to an infinite covering of the Euclidean BTZ black hole \cite{Paper1}. However, here we find that the geometry sourced by the 
presence of a third black hole operator turns out to be a Euclidean 
wormhole with two asymptotic boundaries. 
We construct this new geometry as a quotient of empty AdS$_3$ realized by domes and doors.
The doors give access to the infinite covers that are needed to describe the insertion of the operators, 
while the domes describe the fundamental domains of the quotient on each cover.
In particular, despite the standard fact that the Fefferman-Graham expansion is single-sided, 
the extended bulk geometry contains a wormhole that connects two asymptotic boundaries. 
We observe that the two-sided wormhole can be made single-sided by 
cutting off the wormhole and 
gluing on a ``Lorentzian cap''. In this way, 
the geometry gives the holographic description of a three-point function, up to phases. 
By rewriting the metric in terms of a Liouville field, we 
compute the on-shell action and 
find that the result matches with the Heavy-Heavy-Heavy three-point 
function predicted by the modular bootstrap. 
Finally, we describe the geometric transition between doors and defects, 
that is, when one or more dual operators describe a conical defect insertion, 
rather than a black hole insertion. \\

\end{abstract}

\newpage

\setcounter{page}{1}
\renewcommand{\thefootnote}{\arabic{footnote}}
\setcounter{footnote}{0}



{
\tableofcontents
}

\newpage

\section{Introduction}

In \cite{Paper1}, we initiated the study of holographic 
correlation functions in AdS$_{d+1}$/CFT$_d$ 
that involve huge operators at the insertion points.
By ``huge'', we mean operators of very large conformal dimension, 
such that their dual description is heavy enough to backreact and change the bulk geometry (\ie $\Delta\sim L_\mt{AdS}^{d-1}/\GN$).
For two-point functions, the new backreacted geometries 
were referred to as “spacetime bananas”, and to illustrate 
their features,  we discussed heavy scalar operators that 
create an AdS-Schwarzschild black hole in the bulk.
Inspired by this two-point function construction, 
one of the goals we envisioned in \cite{Paper1} was to
understand what geometries describe three- 
and higher point functions for huge operators.
In this paper, we address this program for 
asymptotically AdS$_3$ spacetimes. By taking advantage 
of the simplicity of pure three-dimensional Einstein gravity, 
we will construct the geometries dual to huge three-point functions 
and establish a general formalism of ``domes and doors" 
which could be applied in the future to higher multipoint correlators.

As discussed in \cite{Paper1}, the backreaction of the dual operators 
on the bulk geometry induces an expectation value for the boundary 
stress tensor \cite{Balasubramanian:1999re}, which is interpreted as
\beq
\langle\, T_{ij}(\vec{\bf x})\,\rangle =
\frac{\Big\langle T_{ij}(\vec{\bf x})\,\prod\limits_{k=1}^n O_{\Delta_k}(\vec{x}_k) \Big\rangle}{\Big\langle \prod\limits_{i=k}^n O_{\Delta_k}(\vec{x}_k)\Big\rangle}\,.
\label{barn7}
\eeq
That is, the boundary expectation value corresponds to
the stress tensor induced by the insertion of the various huge 
operators in the correlation function. We make use of this general 
result with two simplifications in the present work: First, we set 
the boundary dimension $d=2$, \ie we examine holographic correlations 
functions for asymptotically AdS$_3$ geometries. Second, we examine 
the correlation function of three huge scalar operators, \ie $n=3$. 
Let us add that for the most part, our calculations are done in 
Euclidean signature.\footnote{We will consider adding a Lorentzian 
component to the geometry in section \ref{stop}.} 

It will be convenient to parameterize the conformal dimensions of the 
three heavy scalar primary operators as $\Delta_j = \frac{c}{12} M_j$, 
where $c={3L_\mt{AdS}}/{2\GN}$ is the central charge in the boundary 
CFT$_2$. Now taking the two-dimensional boundary geometry to be the 
plane,\footnote{Here $z$ is the usual complex coordinate on the plane.} 
we insert these operators at locations $z_j$ and eq.~\reef{barn7} yields \beq
\langle T_{zz}(z)\rangle = \frac{c}{24}\,\frac{1}{(z-z_1)(z-z_2)(z-z_3)} \sum_{i}\frac{M_i \prod_{j\neq i}(z_i-z_j)}{z-z_i}
\label{L}
\eeq
for the holomorphic component of the stress tensor.
Of course, the anti-holomorphic stress tensor has an identical 
expectation value with $z\to\bar{z}$. 

In passing, we note that for any (order one) value of the $M_j$, 
the insertion of these boundary operators will result in a deformed 
bulk geometry. However, there are two distinct regimes: for $M_j>1$, 
the operators are dual to black holes while for $M_j<1$, they describe 
conical defects in the bulk. For the most part, we focus on the black hole regime.
However, we return to consider defect geometries in section \ref{Defects}.

Now combining data of the flat metric for the CFT$_2$ background and 
the expectation value of the stress tensor, we have sufficient boundary 
conditions to solve for the bulk geometry in a Fefferman-Graham (FG) 
expansion~\cite{deHaro:2000vlm}. A remarkable fact about three-dimensional 
gravity is that the FG expansion of the metric sourced by \eqref{L} truncates 
at a finite order. The resulting metric was studied by Ba\~nados \cite{Banados:1998gg},
\beq
\label{eq:Banados}
ds^2 = \frac{dy^2+dz d\bar{z}}{y^2} + L(z) dz^2 + \bar{L}(\bar{z}) d\bar{z}^2 + y^2 L(z) \bar{L}(\bar{z}) dz d\bar{z}
\eeq
where $\langle T_{zz}(z)\rangle = -\frac{c}{6} L(z)$
and $\langle T_{\bz\bz}(\bz)\rangle = -\frac{c}{6} \bar{L}(\bz)$. This metric is a solution to Einstein's equations for any $L$ and $\bar L$. The simplicity of the FG expansion, which truncates at order $y^2$, can be seen as 
a reflection of the lack of propagating degrees of freedom in three-dimensional Einstein gravity.

\begin{figure}
\centering{
\includegraphics[scale=0.3]{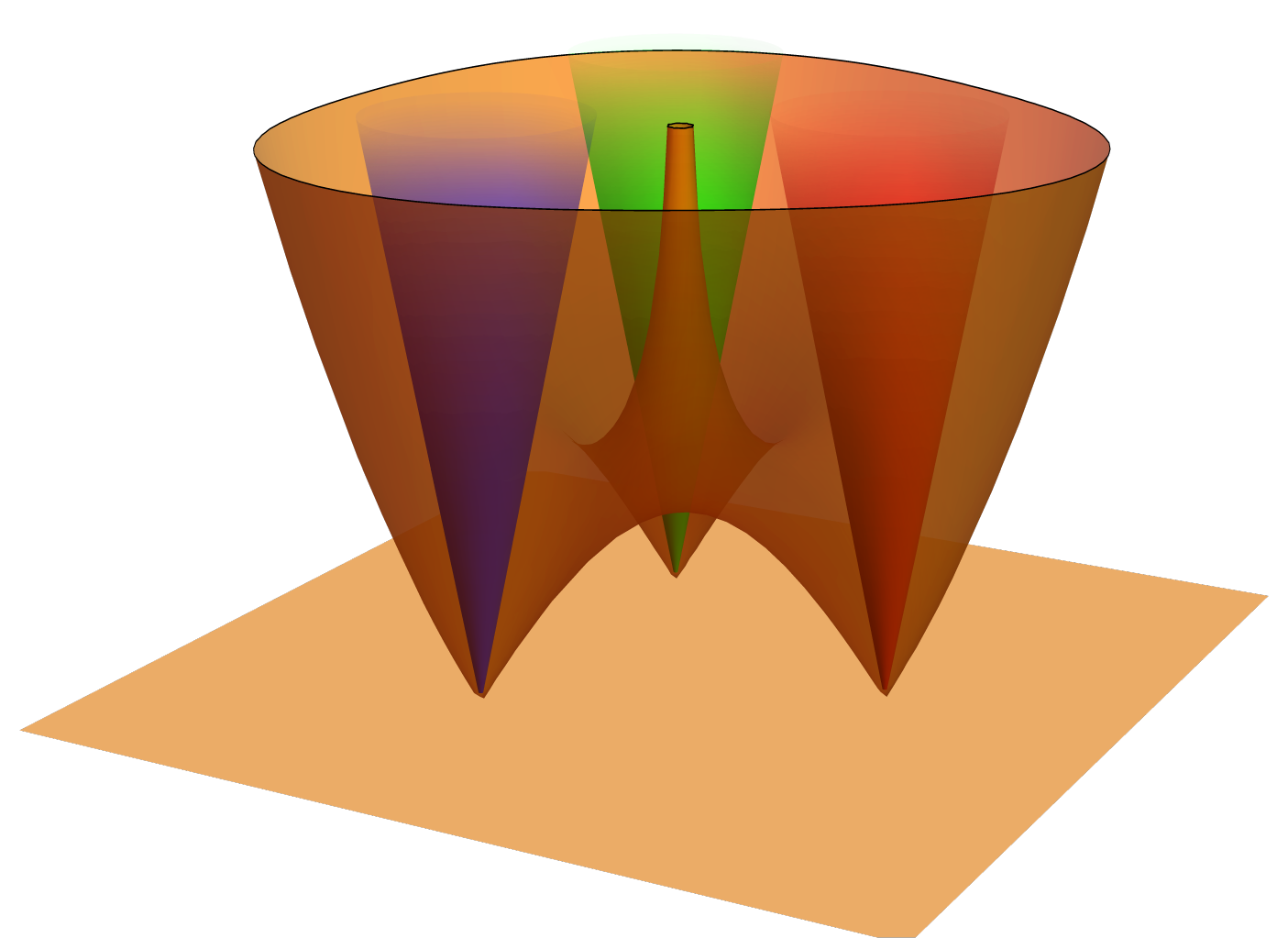}}
\caption{The $\det(g)=0$ wall for the three-point geometry, with the horizons hidden within.} \label{wall}
\end{figure}

Simple as it may appear, the Ba\~nados metric \reef{eq:Banados}
with stress tensor \eqref{L} contains a great deal of interesting physics.
First of all, the coordinates in this metric do not cover the full space: 
there is a coordinate singularity where $\det (g)=0$,  which we call the ``wall'', 
and is depicted in figure \ref{wall}. 
The presence of a wall is a generic of the Ba\~nados metric \reef{eq:Banados}. 
For example, there is also a wall for the two-point function geometry, 
as noted in \cite{Paper1}. However, for the two-point correlator, the 
geometry beyond the wall is easily revealed by a coordinate transformation 
to the familiar ones for the AdS-Schwarzschild black hole \cite{Paper1}. 
For the three-point correlator \eqref{L},  the extension is not immediate 
and needs to be constructed. This will be one of the main goals of this paper. 

One approach of constructing the complete three-point function geometry 
is to exploit the fact that all solutions in three-dimensional Einstein 
gravity with negative cosmological constant are locally isometric to 
AdS$_3$. That is, all solutions can be cast as AdS$_3$ with identifications.
To find the required identifications, we will start from the beautiful 
analysis done by Roberts \cite{Roberts:2012aq} -- see Appendix \ref{RobertsAppendix} for a review -- 
where it was shown that the change of coordinates
\beq
\Big(Y,Z,\bar Z\Big)=\Big(0,f,\bar f\Big)+ \frac{y}{f' \bar{f}' + y^2 \, f'' \,  \bar{f}''/4 }
\Big(( f'  \bar{f}')^{3/2},- y (f')^2 \bar{f}''/2,- y (\bar{f}')^2 f''/2\Big)  \label{Roberts}
\eeq
maps the solution with the Ba\~nados metric (\ref{eq:Banados}) into Euclidean AdS$_3$ in Poincar\'e coordinates,
\beq
ds^2_{{\rm AdS}_3} = \frac{dY^2 + dZ d\bar{Z}}{Y^2}\,. \la{emptyMetric}
\eeq
This map is determined in terms of two boundary functions 
$f(z)$ and $\bar f(\bar z)$, and the Schwarzian derivatives 
of these determine $L(z)$ and $\bar{L}(\bar z)$,
\beq\label{Roberts_schwartz}
L(z)=-\tfrac{1}{2} \{f,z \}\qquad;\qquad \bar{L}(\bar{z})=-\tfrac{1}{2} \{\bar{f},\bar{z} \}.
\eeq
Thus, from \eqref{Roberts}, 
we will be able to understand what identifications of AdS$_3$ are needed in order to construct the three-point geometry,
and in particular, the extension beyond the wall.

\begin{figure}
\centering{
\includegraphics[scale=.57]{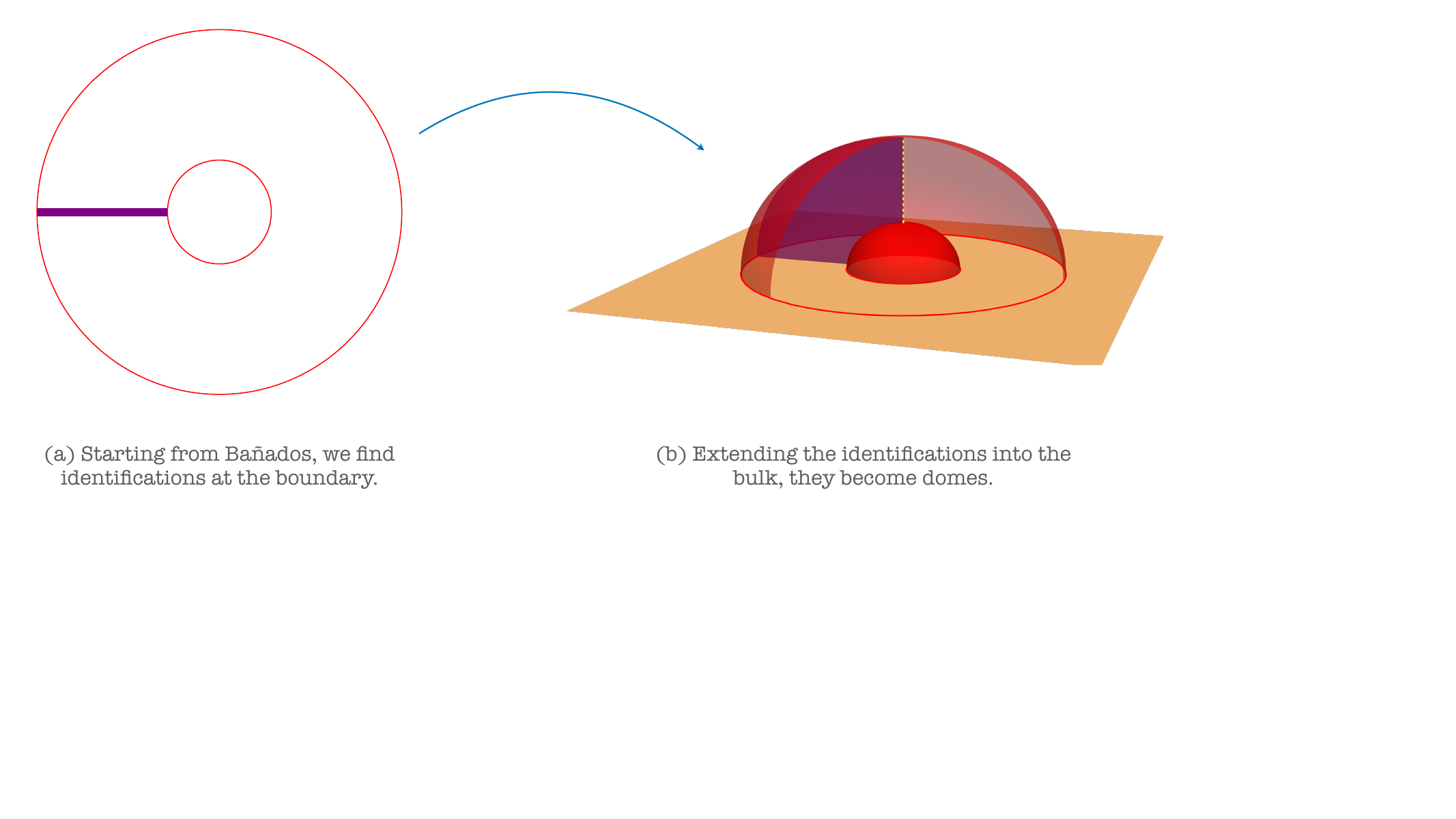}
}
\vspace{-5.2cm}
\caption{Representing the Euclidean BTZ black hole in terms of AdS$_3$ with identifications. On the boundary shown in panel (a), one finds a torus where the two circles are identified. 
Extending the identification into the bulk, these circles become the asymptotic boundaries of the domes shown in panel (b).}
\label{torusFigure}
\end{figure}

Considering the two-point function geometry as an example, 
one can represent the Euclidean BTZ black hole in terms of 
AdS$_3$ with identifications -- see figure \ref{torusFigure}. 
On the boundary, one finds a torus described by identifying (up to a Weyl rescaling) a pair
of circles in the ($Z,\bar{Z})$-plane.
Extending the identification into the bulk, these circles become 
the asymptotic boundaries of two domes, which are identified in the 
AdS$_3$ bulk. Now, the angular direction in the ($Z,\bar{Z})$-plane 
represents the Euclidean time direction. In the standard approach 
of black hole thermodynamics, an angular periodicity of $2\pi$ is 
imposed, which results in a smooth geometry in the bulk, \ie the usual 
hyperbolic solid torus considered in \eg \cite{Krasnov:2000zq,Carlip:1995qv,Maloney:2007ud}.
However, for the two-point correlator \cite{Paper1}, we need to decompactify 
the Euclidean time direction and introduce instead a branch cut 
in the ($Z,\bar{Z})$-plane -- the purple line in figure \ref{torusFigure}(a). 
The figure then represents one sheet of an infinite cylindrical covering geometry. 
In the bulk, this cut extends to a two-dimensional surface, which we denote as 
the ``door'' -- the purple surface in figure \ref{torusFigure}(b). The door 
reaches up to the conical singularity at the horizon (the vertical, dashed line). 
As on the boundary, there is now an infinite sequence of identical sheets 
in the bulk. Approaching one of the black hole operator insertions 
means circling around the geometry and going through the door infinitely many times.

Explaining the corresponding picture for the three-point function is 
one of two key results. The other will be reproducing the universal 
formula for the OPE structure constant in the regime of huge operators, 
as predicted in \cite{Collier:2019weq} (see also \cite{Cardy:2017qhl,Belin:2017nze,Cho:2017fzo}).
This universal OPE formula is related to the structure constants of 
Liouville theory, expressed by the DOZZ formula 
\cite{Dorn:1994xn, Zamolodchikov:1995aa, Teschner:1995yf} (see also \cite{Harlow:2011ny}).
As we will see, the holographic renormalization of the geometries 
dual to correlation functions of heavy operators involves classical 
solutions of the Liouville equations. It is through this connection 
that we find the anticipated universal expression for the OPE coefficients. 

The remainder of the paper is organized as follows: In section 
\ref{Two} we examine description of the two-point function geometry 
in terms of ``domes and doors", in more detail. Of course, this is 
simply a new perspective of the same geometry that we discussed 
in \cite{Paper1}. This discussion is a warm up to familiarize the 
reader with this new description, which we then apply to describe the 
the three point function geometry in section \ref{Three}. 
For three black hole operators, we will see that the 
extension of the Ba\~nados metric involves a single room with 
three doors, and six infinite sequences of rooms with one door each.
In section \ref{Lio}, we explain a connection between these 
three-dimensional solutions and classical solutions in Liouville theory.
Then, in section \ref{Action}, we exploit this connection to 
evaluate the onshell action of these geometries.
In agreement with predictions from the modular bootstrap 
\cite{Collier:2019weq}, we will reproduce the expected classical 
limit of the Liouville three-point function. 
As noted above, there are two regimes for the conformal 
dimensions, producing either black holes or conical defects in 
the bulk. In section \ref{Defects}, we discuss the latter. Although
the defect geometries are considerably different, \eg see figure 
\ref{DOZZ}, we show that the domes-and-doors framework smoothly 
interpolates from black holes to the defect geometries studied in \cite{Chandra:2022bqq, Chang:2016ftb}.
We conclude by listing some open problems and musing about 
some higher dimensional speculations in section \ref{discussion}.

\section{Two-point Function. The Banana is a Door.} \label{Two}

The two-point function case is given by the Ba\~nados metric 
(\ref{eq:Banados}) with\footnote{The resulting Ba\~nados metric 
coincides with the cone metric considered in \cite{Paper1} 
for three-dimensions, namely 
\beq\notag
ds^2_{}=\frac{1}{{\bf y}^2} \left[ d{\bf y}^2 + 
(1-\tfrac{M{\bf y}^2}{4{\bf R}^2})^2  d{\bf R}^2 +   
(1+\tfrac{M{\bf y}^2}{4{\bf R}^2})^{2}{\bf R}^2 d\theta^2\right]
\la{banadosCone}
\eeq
where we introduced polar coordinates on the $z$ plane, $z={\bf R} e^{i \theta}$.
} 
\beq 
L(z)=-\frac{M (z_1-z_2)^2}{4 (z-z_1)^2(z-z_2)^2} \,.
\label{barn8}
\eeq 
Here (and henceforth in this paper) we are setting $\bar L=L$ 
as is the case for spinless objects.
This geometry also arises in the study of the entanglement properties of the state prepared by one of the heavy operators acting on the vacuum, see \cite{Asplund:2014coa}.

\begin{figure}[t]
\centering{
\includegraphics[scale=.4]{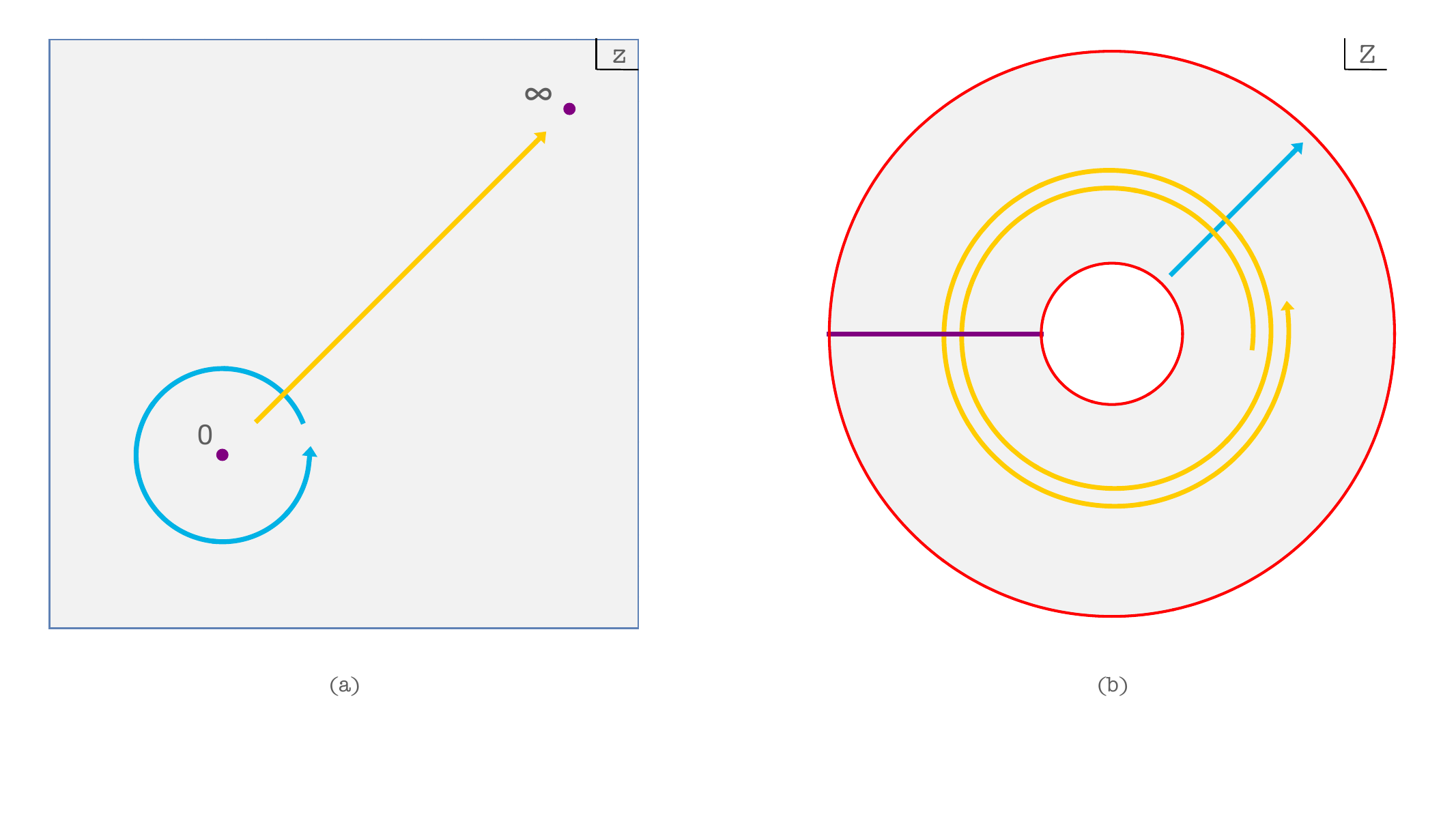}
}
\vspace{-1.5cm}
\caption{The map $z\to Z= z^{i R_h}$ swaps dilatations and rotations in the two planes. The identification $z = e^{2\pi i} z$ restricts the region of interest in the ($Z,\bZ$)-plane to an annulus. Moving along an arbitrary ray in the ($z,\bz$)-plane, from $0$ and $\infty$, translates into the need for an infinite covering of this annulus in the ($Z,\bZ$)-plane. Hence, we introduce a branch cut represented by the purple line on the right. Passing through this cut infinitely many times in the clockwise/anti-clockwise direction means asymptotically approaching the origin/infinity in the ($z,\bz$)-plane, where the two insertions lie.}
\label{doorBoundary}
\end{figure}

As already highlighted in \cite{Paper1}, 
the Ba\~nados metric only describes the two-point
geometry in a coordinate patch which extends from the asymptotic AdS boundary (at small $y$) 
up to where $\det(g)=0$, \ie the surface 
\beq
y^4 L(z) \bar{L}(\bar{z})=1 \,.
\label{barn88}
\eeq
Note that when the boundary stress tensor has poles, as it 
always does when there are local operator insertions in the CFT$_2$, 
this surface will extend all the way to the conformal boundary at $y=0$.

\begin{figure}[t]
\centering{
\includegraphics[scale=.45]{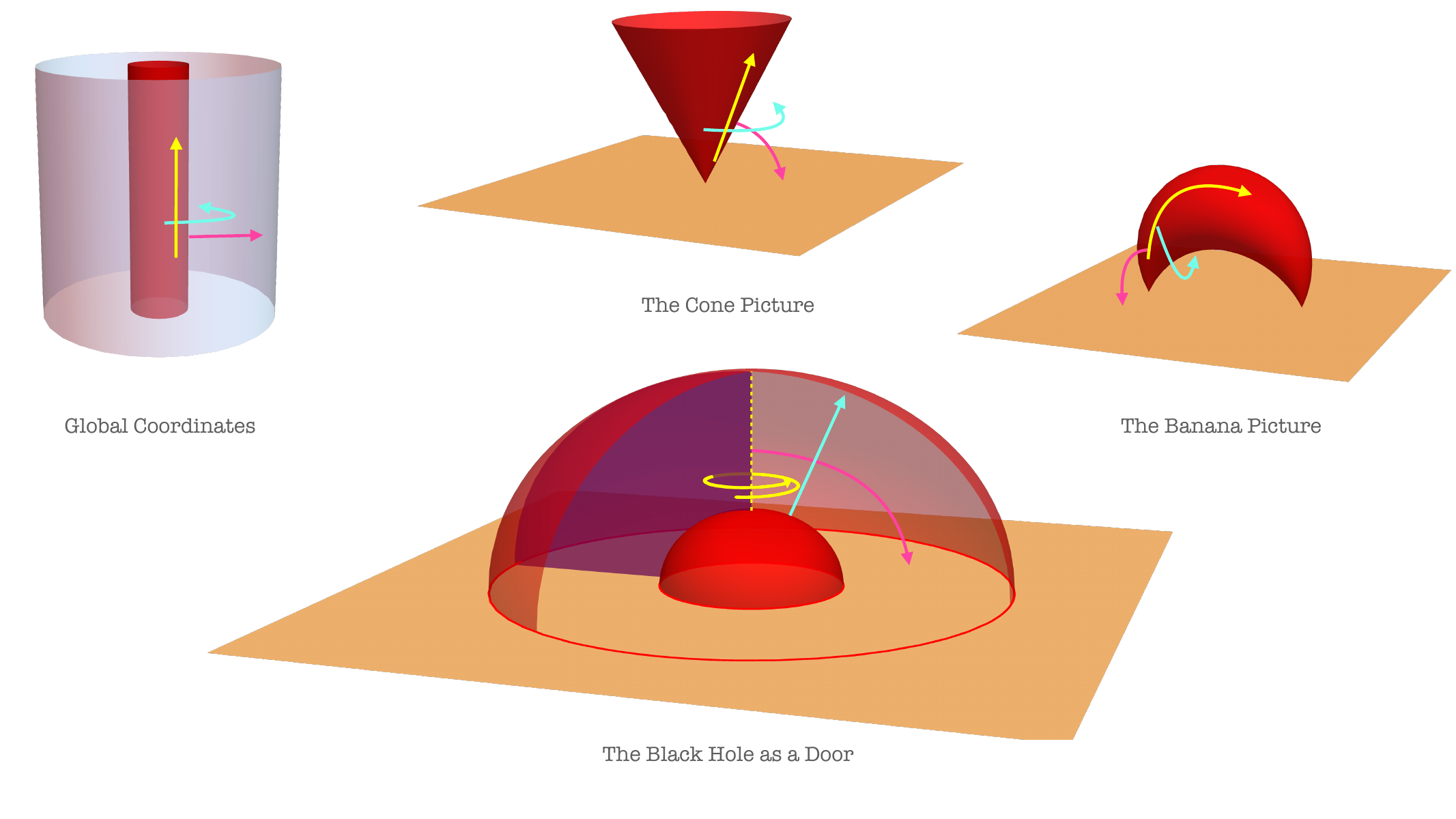}
}
\vspace{-0.5cm}
\caption{In the bottom panel, the two-point function geometry is shown as an infinite branched cover of the solid torus usually used to describe the Euclidean BTZ black hole.  
To represent the covering space, we introduce a branch cut or ``door'', shown in purple. Passing through the door, leads to an identical copy of the solid torus.
The top three panels show the geometry adapted to three different coordinate systems considered in \cite{Paper1}. The far left panel shows the directions for the Euclidean time (yellow), radius (red) and angle (green) in global coordinates. The corresponding directions are shown in the other panels for other coordinate systems. In the dome-and-door illustration, the Euclidean time direction corresponds to angular rotations in the ($Z,\bZ$)-plane. Hence, approaching one of the operator insertions corresponds to rotating infinitely many times in one direction or the other. 
The axis of this rotation is the horizon (the vertical yellow dashed line).}
\label{bulkDoor}
\end{figure}

As noted in the introduction, we can extend beyond the 
wall by finding the identifications needed to construct 
the full geometry as a quotient of Euclidean AdS$_3$.
These identifications can be read off from the Roberts map (\ref{Roberts}) by noting 
that the boundary $y=0$ and $(z,\bar z)$ is mapped into the boundary $Y=0$ with $(Z,\bar Z)=(f(z),\bar f(\bar z))$.
Hence these two functions have a clean geometrical interpretation as describing how the two boundary geometries map into one another. 
Similar considerations can be drawn for a neighbourhood of the boundary, where the Roberts map is well-defined. 

For a  two-point function in the black hole regime (\ie with $M>1$), with the insertion points at zero and infinity,  eq.~\reef{barn8} reduces to $L(z)=-\frac{M}{4 z^2}$ and $\bar L(\bar z)=-\frac{M}{4 \bz^2}$, and we find
\beq
f_\texttt{BH}(z)=z^{iR_h} \qquad;\qquad \bar{f}_\texttt{BH}(\bar{z})=\bar{z}^{-iR_h}\qquad;\qquad R_h=\sqrt{M-1}\in\mathbb{R}^+\ .
\label{barn9}
\eeq
Recall that $R_h = \sqrt{M -1}$  would correspond to the horizon radius of the black hole in global coordinates. 
Note that $f_\texttt{BH}$ and $\bar{f}_\texttt{BH}$ are conjugates of each other, and so the Roberts map 
sends the Ba\~nados patch into the real Euclidean AdS$_3$.
Because the exponents are purely imaginary, we see that travelling along a ray from the origin to infinity in the $z$ plane will correspond to rotating infinitely many times in the ($Z,\bZ$)-plane around the origin. That is, on the boundary, eq.~\reef{barn9} yields
\begin{equation}\label{radially_to_door}
z=e^{\tau+i\phi}
\quad\longrightarrow\quad\ Z=  e^{iR_h\tau - R_h\phi}
\end{equation}
Hence going to $z=0$ at fixed $\phi$ corresponds to $\tau\to -\infty$, which corresponds to an infinite number of clockwise rotations at fixed radius in the the ($Z,\bZ$)-plane, and similarly for $z\to\infty$. 
In order to cover the two-point function geometry
we need to cover the $(Z,\bar{Z})$-plane infinitely many times as we go around the origin. To connect all these infinitely many (in this case) identical sheets, we add a branch cut anchored at the origin and extending out along the negative real axis (\ie $\phi=\pi$). Crossing the cut allows one to move from one sheet to another on the boundary. 

Conversely, rotating around the $(z,\bar z)=(0,0)$ point corresponds to travelling along a ray in the $(Z,\bar Z)$-plane.
Under a $2\pi$ rotation in the $z$ plane, we find $\sqrt{Z \bar Z} \to \sqrt{Z \bar Z} e^{2\pi R_h}$. Hence we can restrict region of interest in the $(Z,\bar Z)$ plane to an annulus between two radii, $R_1$ and $R_2=R_1 e^{2\pi R_h}$, which are identified. Any choice of $R_1$ is as good as any other since this is just a choice of origin for the polar angle in the $(z,\bar{z})$-plane. 
All in all, we end up with the geometry described in figure \ref{doorBoundary}.

\begin{figure}[t]
\centering{
\includegraphics[scale=.47]{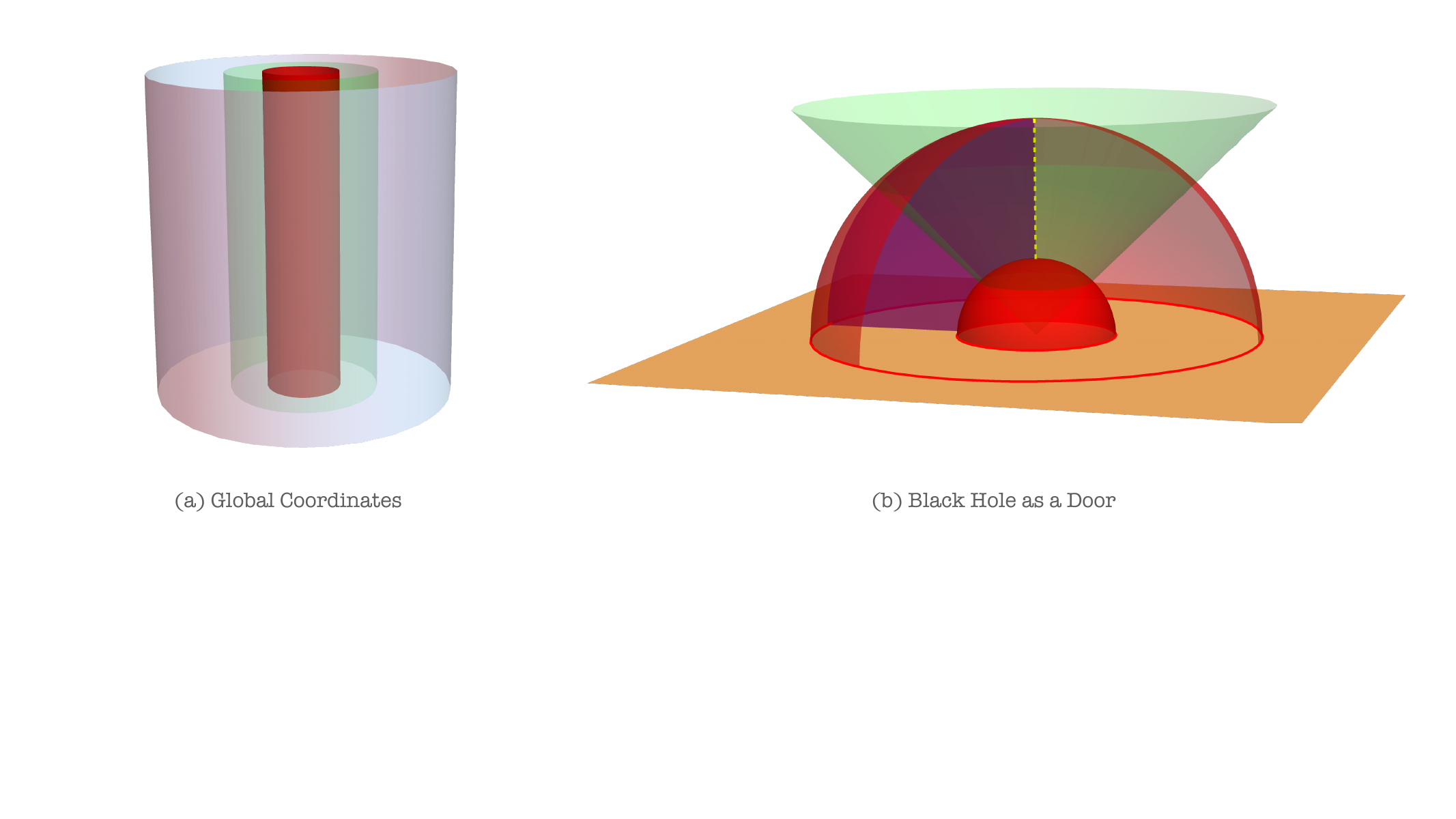}
}
\vspace{-4cm}
\caption{In the Ba\~nados patch, the horizon is hidden behind the det($g$) wall (in green).  In global coordinates in panel (a), the horizon is at $r=R_h=\sqrt{M-1}$ while the wall sits at $r=\sqrt{M}>R_h$.  Under the Roberts map, the det($g$) wall is mapped to a cone shown in panel~(b).
}
\label{coneExtension}
\end{figure}

The boundary identifications and branch cuts in the ($Z,\bZ$)-plane extend into the bulk, as shown in the bottom panel of  figure \ref{bulkDoor}. Extending the cut to a two-dimensional surface in the bulk leads to our notion of a ``door'' (shown in purple). Note that in the bulk, the branch cut is also bounded by the axis of rotation, \ie the vertical line $Z=0=\bZ$. In passing through the door, one moves from one sheet to another in the bulk. Similarly, the boundary annulus 
is promoted to the bulk region between a pair of concentric domes. These domes are identified and so upon passing through the outer dome, one emerges again into the same region from the inner dome. The black hole horizon is the geodesic line $(Z,\bar Z)=0$ that travels vertically between the domes, and as noted above, it serves as part of the doorframe.

As emphasized in figure \ref{bulkDoor}, the dome-and-door construction is simply an alternative description of the geometry studied in \cite{Paper1}. 
In particular, we can write the change
of variables from empty AdS$_3$ in eq.~\eqref{emptyMetric} 
to the ``cone'' metric, by composing the 
known change of variables to the BTZ black hole \cite{Carlip:1994gc}, namely
\beq
Y=\frac{R_h}{r}e^{R_h\phi}\quad;\quad 
Z=\sqrt{1-\frac{R_h^2}{r^2} }\, e^{R_h(\phi+i\tau)}\quad;\quad
\bar{Z}=\sqrt{1-\frac{R_h^2}{r^2}}\, e^{R_h(\phi-i\tau)}\,,
\eeq
with the Global-to-Poincare map introduced in \cite{Paper1}.
The ``banana'' geometry is then obtained by a change of coordinates that acts as a special conformal 
transformation on the boundary. 

Let us add that in the dome-and-door construction, 
we could use isometries of AdS$_3$ to obtain an infinite family of equivalent pictures where
the boundary circles are no longer concentric -- see details in appendix \ref{RobertsAppendix}. In this case, the horizon becomes a semicircular geodesic between the domes. The door again extends between the domes, and from the asymptotic boundary to the horizon. These different realizations of the two-point function geometry will make an appearance in the following section.

To summarize, we note that the Ba\~nados coordinates break down at the wall \reef{barn88}. For the two-point geometry, this becomes the cone shown in green in figure \ref{coneExtension}(b).
This cone corresponds to a cylinder in the global coordinates describing the BTZ black hole, which hides
the horizon behind it, as depicted in figure \ref{coneExtension}(a).
By mapping the Ba\~nados patch isometrically into Euclidean AdS$_3$, we found a simple construction of the solution using identifications along spherical domes plus a door extending between them.
This geometry can be readily extended beyond the wall (\ie the green 
cone) to get the complete geometry depicted in figure \ref{coneExtension}(b).
In this complete geometry, the horizon is the vertical geodesic going from the tip of the inner dome to the tip of the outer dome.
As discussed in the introduction, were it not for the door, this geometry would be the well-known story for how to construct Euclidean AdS-Schwarzschild in three dimensions as a quotient of Euclidean AdS$_3$.

\section{The 3pt Function Wormhole. A Room with Three Doors.} \label{Three}

Having reviewed the two-point function geometry as AdS$_3$ with identifications, we are ready to discuss the three-point function geometry. Recall that the corresponding expectation value of the stress tensor is given by (\ref{L}), namely
\beq
L(z)=-\frac{1}{(z-z_1)(z-z_2)(z-z_3)} \sum_{i}\frac{M_i \prod_{j\neq i}z_{ij}}{4(z-z_i)}
\label{barn99}
\eeq
with $z_{ij}=z_i-z_j$.
Assuming the
operators are inserted at $z_1=0$, $z_2=\infty$ and $z_3=1$ for convenience, 
a particular $f(z)$ that solves the Schwarzian equation \eqref{Roberts_schwartz} 
for $L(z)$ in \eqref{barn99}, and determines the Roberts map \eqref{Roberts}, is
\footnote{Here $\mathcal{N}^2 = 
\frac{\Gamma^2(-i R_1)}{\Gamma^2(+i R_1)} 
\frac{\gamma\left(\frac{1+i (R_1 +R_2+R_3)}{2}\right) 
\gamma\left(\frac{1+i (R_1 -R_2+R_3)}{2}\right)}{
\gamma\left(\frac{1-i (R_1 +R_2-R_3)}{2}\right) 
\gamma\left(\frac{1-i (R_1 -R_2-R_3)}{2}\right)}$, where $\gamma(x) = \Gamma(1-x)/\Gamma(x)$.
Different solutions to the Schwarzian equation are related to each other by M\"obius transformations of $f$, as described in appendix \ref{RobertsAppendix}.
We have made a specific choice here (particularly, for the phase described by $\cal N$), so that the monodromy of $f_{\texttt{3pt}}$ around the operator insertion at z=1 acts as an isometry of the hyperbolic plane, \eg see \cite{Hadasz:2003he}. This choice will ensure that it yields a smooth, single-valued solution to the Liouville equation in section \ref{Lio}.}
\beq
f_\texttt{3pt}(z) = i \mathcal{N} z^{i R_1} \frac{_2 F_1 \left(\frac{1}{2} + i \frac{R_1 - R_2 - R_3}{2}, \frac{1}{2} + i \frac{R_1 + R_2 - R_3}{2};1 + i R_1; z\right)}{_2 F_1 \left(\frac{1}{2} - i \frac{R_1 - R_2 + R_3}{2}, \frac{1}{2} - i \frac{R_1 + R_2 + R_3}{2};1 - i R_1; z\right)}\,.
\label{arc12}
\eeq
Then, we pick $\bar f_\texttt{3pt}$ to be the complex conjugate of $f_\texttt{3pt}$, so that eq.~\eqref{Roberts} is real.
We have introduced $R_i = \sqrt{M_i -1}\in\mathbb{R}^+$, which correspond to the horizon radii for the dual black holes. 

The function $f_\texttt{3pt}(z)$, also known as the \textit{Schwarz triangle function}, 
has branch points at the locations of the operator insertions.
We can choose the branch cuts to run from $0$ to $\infty$ along the 
negative real $z$ axis, and from $1$ to $\infty$ along the positive real axis.
With this choice, we can then investigate the images of each side of these branch cuts, which is a pair of circles, 
and start building up our 
three-point function geometry.

\begin{figure}
\centering{
\includegraphics[scale=0.4]{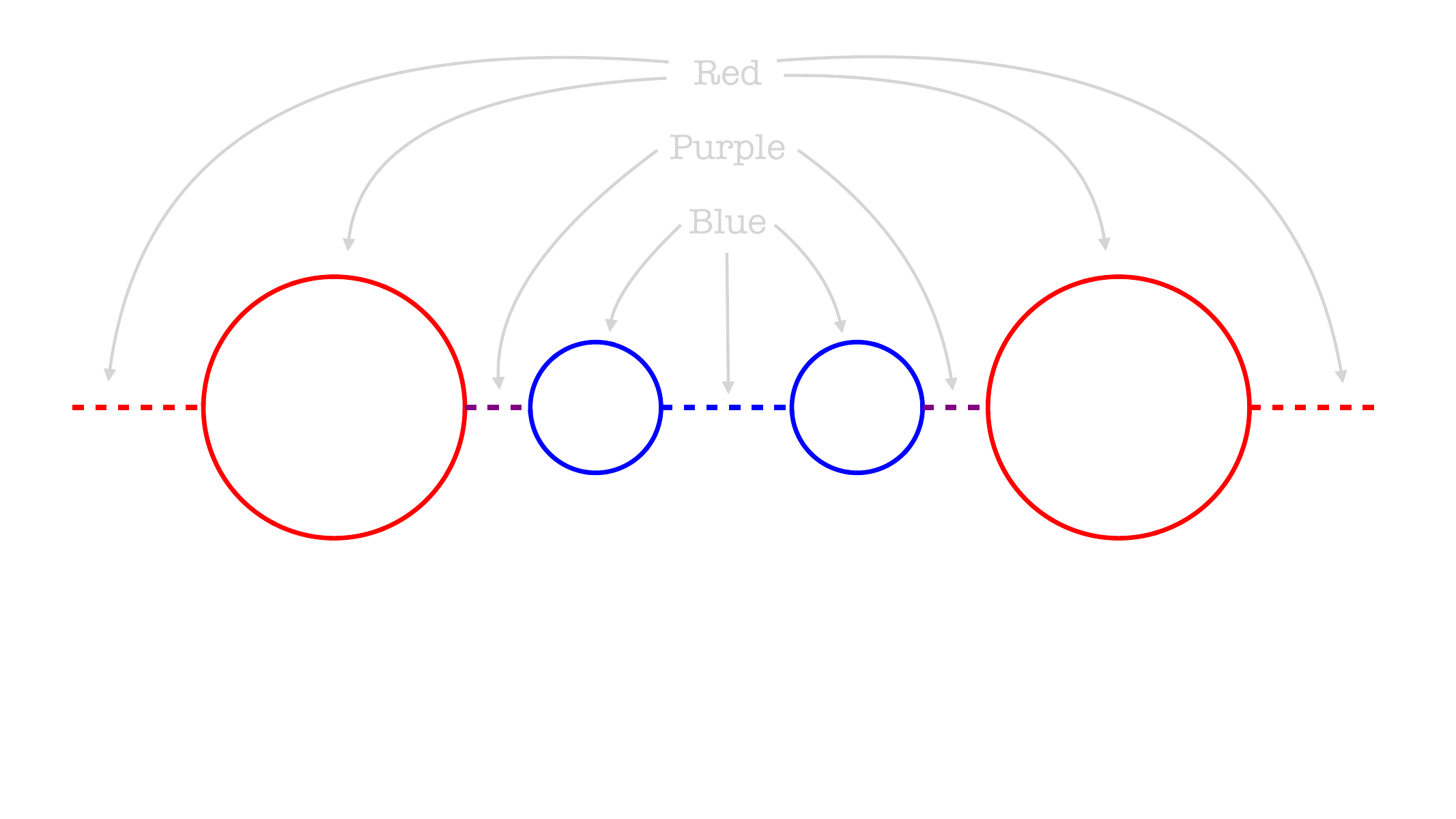}
}
\vspace{-3cm}
\caption{The identifications (solid lines) at the boundary of the three-point function geometry. 
Gluing the circles that share a color creates a genus-two surface. The boundary of the three-point function geometry involves a covering of this surface with infinitely many sheets. To pass between these sheets we introduced a series of branch cuts along the real axis (dashed lines).}
\label{3ptIdents2dOnline}
\end{figure}

In order to describe our geometry in a simple way, we will consider the setup of figure \ref{3ptIdents2dOnline}. 
For this figure we have used the isometries of Euclidean AdS$_3$ to put the circles 
defining the identifications on the boundary into a standard configuration, where they are aligned symmetrically along the real axis.\footnote{This requires composing $f_\texttt{3pt}(z)$ with a M\"obius transformation of the AdS$_3$ boundary -- see appendix \ref{RobertsAppendix}.}
The color code indicates which circles are to be glued to one another. 
Performing these identifications produces a genus-2 surface. 
However, just as in the two-point function geometry, approaching 
one of the operator insertions translates into rotating infinitely 
many times around these circles. The corresponding branch cuts are drawn as dashed lines along the real axis in the figure.\footnote{Note that the purple cut which extends between the red and blue circles has two components that are glued into a single cut by the quotient.}
To have a concrete realization in mind, let us say that 
crossing the red cut infinitely many times
leads to the operator at $z_1$, while crossing the purple door leads to the operator at $z_2$, and crossing the blue door leads to the operator at $z_3$.
Recall that passing through each branch cut leads to a new sheet on the boundary.
The crucial novelty compared to the two-point function case is to realize that on each of these other sheets, we encounter only a subset of circles.
This can be understood by noting that only one of the branch cuts is visible in a domain close to a given operator. Further, let us note that since the cuts cover the entire real axis (between the circles), it is impossible to find a path (in the boundary) which would take us from the bottom of the figure to the top side.

When we extend the boundary identifications into the bulk, each pair of boundary circles extends to a pair of domes which are identified in the bulk.
Further,  there are three (closed) geodesics connecting the domes, which we refer to as the three horizons. The three boundary cuts now extend to three
doors in the bulk, each of which stretch from the asymptotic boundary to the corresponding horizon.
If it were not for the doors, this would be the familiar genus-2 handlebody solution, encountered in \eg \cite{Krasnov:2000zq, Faulkner:2013yia, Balasubramanian:2014hda}.

\begin{figure}
\centering{
\includegraphics[scale=0.45]{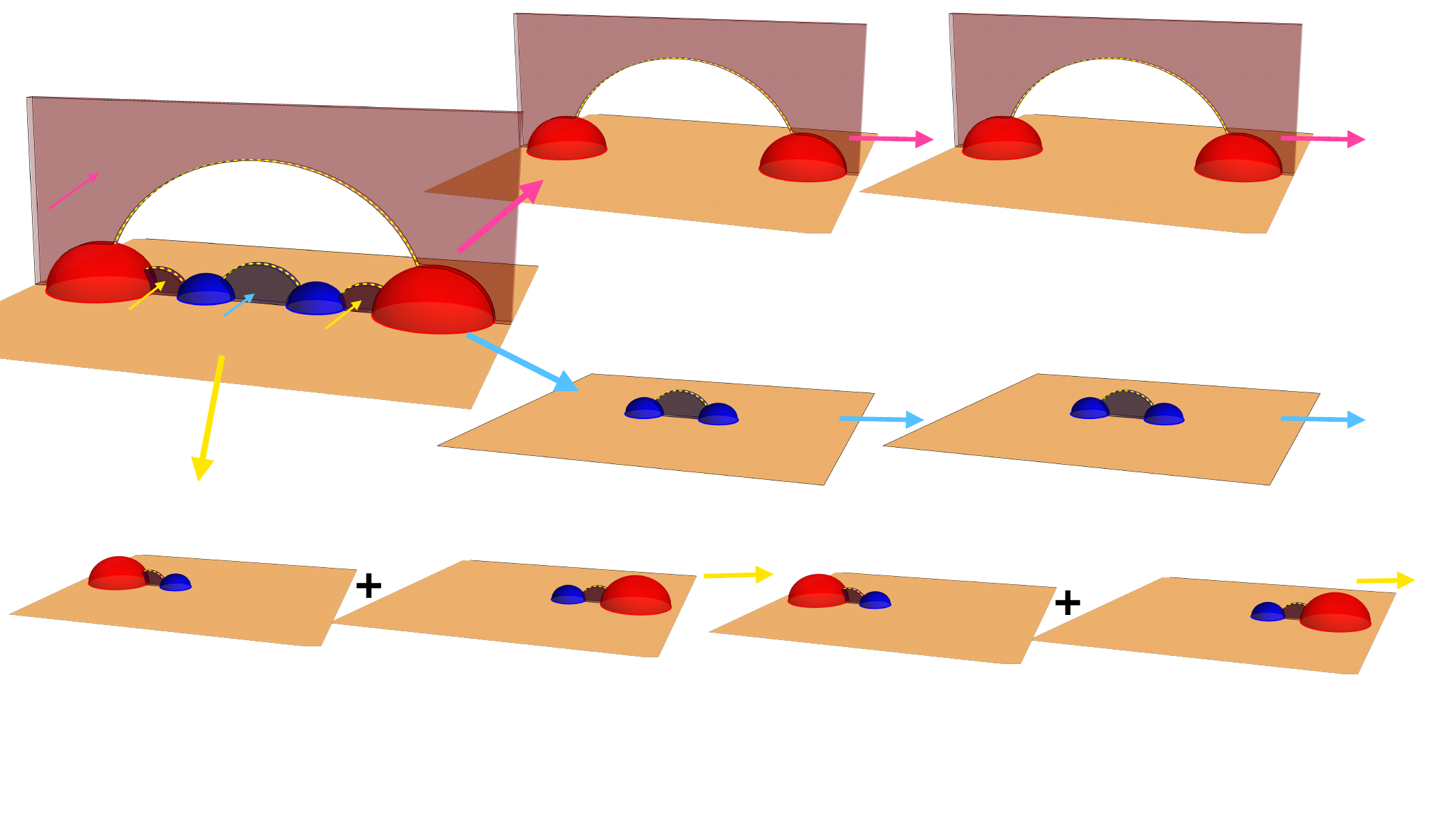}
}
\vspace{-1.5cm}
\caption{The three-point function geometry consists of a central room with all of the domes present and  three doors anchored to the three horizons. Passing through any of these leads to an infinite sequence of rooms with fewer domes. One of the doors (shown in purple) is split into two pieces that are glued together by the domes. This splitting continues as we travel down the corresponding infinite sequence of rooms.}
\label{3ptDoors}
\end{figure}

\subsection{Appearance of a Wormhole} \label{WormH}

From our discussion so far, we have found that the three-point function geometry 
consists of a junction of ``chambers'' -- see figure \ref{3ptDoors}.
There is a single central chamber, which is special because all of the domes and doors appear there.
For each of the three horizons, there is a door and 
passing from the central chamber through one of the doors leads to  a ``leg'', which is sequence of new chambers, each of which is
isometric to the two-point function geometry.
That is, in each of the leg chambers, we see only a subset of the domes and a single door, exactly what is needed to construct the two-point geometry discussed in the previous section. 
So the most surprising aspect of our dome-and-door construction is that when we take into account that we can leave the central chamber through either side of the door, there are in total {\it six} ``legs''.
But the two-point function discussion suggests 
that each leg corresponds to an operator insertion, so we would have expected only  three!

The resolution to this ``puzzle" is found by reconsidering the boundary of the geometry.
As we pointed out above,  the presence of the branch cuts along the real axis (in figure \ref{3ptIdents2dOnline}) splits the boundary of the central chamber into two disconnected pieces. That is, if we start near the bottom of the figure we can only access three of the legs since we can only pass through the cuts from below. However, as shown in figure \ref{3ptDoors},
we can start from one side of the boundary and reach the other by following a path through the bulk that passes through the gap between the three horizons. Any curve going from one side of a given door to the other necessarily 
leaves the boundary and crosses through the bulk.    
So we see that the three-point function geometry 
is not a single sided geometry but rather 
a Euclidean {\it wormhole} with two asymptotic regions!

Hence the central chamber plays a special role in connecting the two asymptotic boundaries of the wormhole. In this context, the ``convex hull" bounded by the three horizons on the surface $Z=\bZ$ becomes the mouth of the wormhole, \ie the extremal surface at the center of the wormhole. A convenient set of coordinates to describe the wormhole metric is given the following ``book page" ansatz \cite{Emparan:1999pm,Maldacena:2004rf},
\beq\label{ini_bookpage}
ds^2 = d\rho^2 + \cosh^2\!\rho\ d\Sigma^2
\eeq
where the metric $d\Sigma^2(w,\bar{w})$ has constant negative curvature.
The two asymptotic AdS$_3$ boundaries are reached with $\rho\rightarrow\pm\infty$.
As we explained above, they are separated by the doors. It will be the 
purpose of section \ref{Lio} to describe $d\Sigma^2$ for the three-point function geometry.

Multi-boundary Euclidean 
geometries have been the subject of many interesting developments, see  \eg \cite{Maldacena:2004rf,Saad:2019lba, Chandra:2022bqq, Cotler:2020ugk, Schlenker:2022dyo, DiUbaldo:2023qli}.
At least in some cases, these geometries can be understood holographically
as contributing to the moments of CFT observables averaged over 
appropriately defined \emph{ensembles} of CFT's.
This suggests that rather than calculating the value of a 
specific three-point function, the geometry we have constructed here 
should be interpreted as a contribution to the \emph{variance} 
of the three-point function over some ensemble of CFT$_2$'s.

\begin{figure}
    \centering
    \includegraphics[scale=0.45]{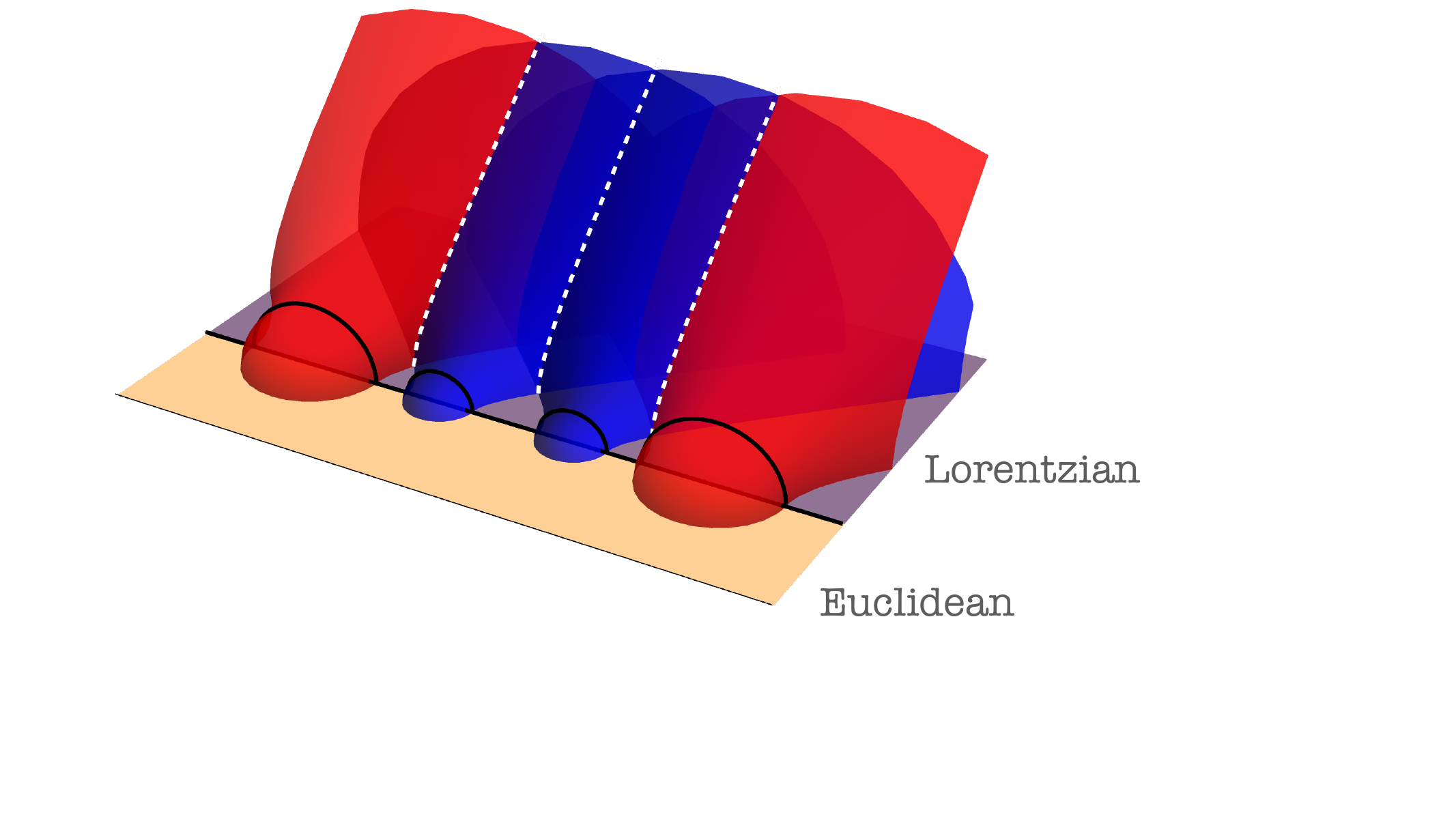}
    \vspace{-2.5cm}
    \caption{ 
    A Lorentzian cap is obtained by Wick rotating $\rho\rightarrow i t$. The 
    domes in the Euclidean part of the geometry, become hyperboloids in Lorentzian  signature. 
    The black hole singularities are the result of the intersections of the hyperboloids, which we represent here as a white dashed lines. Note that there is another singularity `at infinity' where the two red hyperboloids intersect. Further, there is a single asymptotic region between the red and blue hyperboloids, which is split into two pieces in the diagram. (This is analogous to the splitting of the purple door in figure \ref{3ptDoors}.)
    }
    \label{signatureChange}
\end{figure}

\subsection{A Lorentzian Alternative} \label{stop}

We would like to note that there is another approach to extending our dome-and-door construction away from the boundary that results in a single-sided solution. This requires us to continue the geometry to Lorentzian signature at the mouth of the wormhole,
and the effect is to remove one of the two asymptotic boundaries.\footnote{We thank Juan Maldacena for bringing this possibility to our attention.}
This can be achieved by Wick rotating the 
book-page metric \eqref{ini_bookpage} at $\rho = 0$. With $\rho\rightarrow it$, the metric takes a cosmological Friedmann–Robertson–Walker form,
\beq\label{new-bookpage}
ds^2 = -dt^2 + \cos^2t\ d\Sigma^2\,.
\eeq
If we continue the entire geometry beyond $t=0$, the Lorentzian geometry corresponds to (the future half of) a three-sided black hole geometry, \eg see \cite{Aminneborg:1997pz, Brill:1995jv, Skenderis:2009ju,Balasubramanian:2014hda}.
The Euclidean domes that define the identification are analytically continued into hyperboloids in the Lorentzian geometry. 
There are three exterior regions, each of which is isometric to the BTZ black hole exterior, and there is non-trivial spacetime region hidden behind the horizons.
For our purposes, we only perform the Wick rotation at the mouth of the wormhole, and so the Lorentzian geometry only includes a portion of this region hidden behind the horizons. While this leaves our construction with some ambiguity (see further discussion in section \ref{Action}), this is sufficient to remove access to the second asymptotic boundary. 

Let us foreshadow the result of computing the action of these geometries in section \ref{Action}, to understand better the relationship between the wormhole and the single-sided, partially Lorentzian solutions. Since we are dividing the Euclidean wormhole geometry in half, the contribution of the Euclidean portion of the single-sided geometry is exactly half the action of the wormhole. Further,
the action of the Lorentzian region is imaginary and so just contributes a phase to the three-point function.  Overall, we will find 
\beq
I_{\text{single-sided}} = \frac{1}{2} I_{\text{wormhole}}+ i \sum_i f(i)\,,
\eeq
where it is a nontrivial result that the imaginary term takes the form of the sum shown above. This form allows the corresponding
phases to be absorbed into the definition of the operators. \\

\section{Relationship with Liouville Theory}\label{Lio}

We have described the three-point geometry with a variety of different 
coordinate systems and metrics. In particular, we first considered the Ba\~nados 
form \eqref{eq:Banados}, then we used identifications within Poincar\'e 
coordinates \eqref{emptyMetric} to construct the full bulk geometry, and we connected these two descriptions. 
Above, we also introduced the ``book page" ansatz \eqref{ini_bookpage}. 
The primary motivation for the latter there was that it facilitates the 
Lorentzian construction described in section \ref{stop}. Here, we reconsider 
the book-page metric and introduce a description of the three-point geometry 
in terms of a Liouville field \cite{Krasnov:2000zq,Takhtajan:2002cc},
which will be convenient to compute the on-shell action in the next section \ref{Action}, since as it will turn out, the on-shell action reproduces a universal  formula \cite{Collier:2019weq} 
for the OPE structure constants expressed as the classical limit of the 
DOZZ formula \cite{Dorn:1994xn, Zamolodchikov:1995aa, Teschner:1995yf}.

Recall the book-page metric \eqref{ini_bookpage} takes the form \cite{Emparan:1999pm,Maldacena:2004rf}
\beq
ds^2 = d\rho^2 + \cosh^2\!\rho\ d\Sigma^2\,,
\label{dino1}
\eeq
where $d\Sigma^2$ describes a slice of the geometry with constant negative curvature, \ie the book page. Irrespective of the details of these slices, the metric \reef{dino1} is a solution of the three-dimensional Einstein equations with a negative cosmological constant.  

It is useful to see 
the book-page description of empty AdS$_3$, which is simply an AdS$_2$ foliation. 
Consider AdS$_3$ described by Ba\~nados metric \eqref{eq:Banados} with $L(z)=0=\bar{L}(\bar{z})$ or the Poincar\'e metric \eqref{emptyMetric}, the change of variables to eq.~\reef{dino1} is
\beq
y=Y=\frac{w_1}{\cosh\rho}\qquad{\rm and}\qquad z=Z= w_1\tanh(\rho)+i w_2\,,
\label{change78}
\eeq
with $w=w_1+iw_2$. Eq.~\reef{change78} then yields
\beq
ds^2 = d\rho^2 + \cosh^2\!\rho\ \frac{4\,dw\, d\bar{w}}{(w + \bar{w})^2} \,.
\label{H3BookPages}
\eeq
Note that fixing $w_1$ and $w_2$ to be constants, 
the paths described by $\rho$ running over its full range are semicircular geodesics 
anchored at the boundary at $z=Z=\pm w_1+iw_2$. As noted previously, we reach the asymptotic boundary (\ie $y=0=Y$) with $\rho\to\pm\infty$. However, we can also reach the boundary on any fixed-$\rho$ slice by taking $w_1=\text{Re}(w)\to0$. That is, all of the book pages (\ie two-dimensional slices) meet the asymptotic boundary along the imaginary axis, $\text{Re}(z)=0=\text{Re}(Z)$.\footnote{Hence, we see the complete ``book" of the book-page ansatz \reef{dino1}. The front and back covers of the book lie flat on the asymptotic boundary $y=0$, and are reached with $\rho\to\pm\infty$, respectively. Of course, the two-dimensional AdS$_2$ slices foliating the bulk geometry correspond to the book pages. Lastly, these pages all meet at the spine of the book at $y=0$, $\Re(z)=0$, which is reached with $w_1\to0$. }

Clearly in the above example, where $\Sigma$ is noncompact, the corresponding solution \eqref{H3BookPages} has a single asymptotic region. More generally, the solutions \reef{dino1}  have two separate boundaries at $\rho\to\pm\infty$.
This is clear when $\Sigma$ is a compact (negatively-curved) surface, which yields  a two-sided wormhole where the two asymptotic regions have the topology of $\Sigma$ \cite{Maldacena:2004rf}. 

While it is straightforward to extend the ansatz \reef{dino1} to any number of dimensions \cite{Emparan:1999pm,Maldacena:2004rf}, there is a particular result which is special to three bulk dimensions. In particular with $d=2$, we observe that the line element $d\Sigma^2$ can be described by a field $\varphi(w,\bar{w})$ 
that satisfies the classical Liouville equation with a negative cosmological constant.  That is,
\beq %
d\Sigma^2 = e^{\varphi(w,\bar{w})}\, dw\, d\bar{w}\qquad{\rm where}\quad
\partial \bar{\partial}\,\varphi(w,\bar{w}) = \frac{1}{2} e^{\varphi(w,\bar{w})}\,.
\label{Kentucky}
\eeq
This allows us to describe our multi-point geometries for huge operators in terms of solutions of the Liouville equation.
To proceed, we only need to determine the appropriate Liouville solution.
This can be done by working out locally the change of coordinates 
to the asymptotic FG patch.
This change of coordinates involves the three functions $\rho=\rho(y,z,\bar{z})$, 
$w=w(y,z,\bar{z})$ and $\bar{w}=\bar{w}(y,z,\bar{z})$,
such that in a small $y$ expansion the book page metric 
is brought into the Ba\~nados form with given $L(z),\bar{L}(\bar{z})$. 
This then yields the relation between the Liouville field $\varphi(z,\bar{z})$ 
and the profile of the stress tensor $L(z),\bar{L}(\bar{z})$.
The first few terms of the FG expansion are
\begin{align}
\rho=&-\log(y/2) + \tfrac{1}{2}\varphi(z,\bar{z})
+ \frac{y^2}{4}\,\partial\varphi(z,\bar{z})\,\bar{\partial}\varphi(z,\bar{z})+ \cdots\,,\\
w=&\,z- \frac{y^2}{2}\,\partial\varphi(z,\bar{z}) +\cdots\,, 
\qquad\quad
\bar{w}=\,\bar{z}- \frac{y^2}{2}\,\bar\partial\varphi(z,\bar{z}) +\cdots\,.
\label{lio_to_FG}
\end{align}
Note that the above change of variables is an infinite expansion in $y$, 
but once we substitute into the metric, the metric truncates at order $y^2$ giving the  Ba\~nados metric.
Then, the relation between the Liouville field $\varphi(z,\bar{z})$ and the stress tensor $L(z),\bar{L}(\bar{z})$ is\footnote{It is straightforward to confirm that these expression yield $\bar\partial L=0=\partial\bar L$, as desired, using the Liouville equation of motion \reef{Kentucky}.}
\beqa
\label{phi_toL}
L(z)&=&\frac{1}{4}(\partial\varphi(z,\bar{z}))^2-\frac{1}{2}\,\partial^2\varphi(z,\bar{z})\,, \\
\bar{L}(\bz)&=&\frac{1}{4}(\bar{\partial}\varphi(z,\bar{z}))^2-\frac{1}{2}\,\bar\partial^2\varphi(z,\bar{z})\,.
\nonumber
\eeqa
The general solution to the Liouville equation \eqref{Kentucky} can be parameterized by a holomorphic and an anti-holomorphic function, $f$ and $\bar{f}$, with
\beq
\label{eq:liouvilleFromQuotient}
\varphi(z,\bar{z}) = \ln\left(\frac{4 
\,\partial f\,\bar{\partial}\bar{f}}{\left( f + \bar{f} \,\right)^2}\right)\eeq
Substituting this expression into eq.~\eqref{phi_toL}, we find 
\beq
L(z)=-\tfrac{1}{2} \{f,z \} \qquad;\qquad 
\bar{L}(\bar{z})=-\tfrac{1}{2} \{\bar{f},\bar{z} \}
\eeq
Hence, we have recovered the Schwarzian equations \eqref{Roberts_schwartz} 
appearing in the discussion of the Roberts map. That is,  $f(z)$ 
and $\bar{f}(\bar{z})$ in eq.~\reef{eq:liouvilleFromQuotient} are the same functions
appearing in the Roberts change of variables \eqref{Roberts} 
from Euclidean AdS$_3$ to the Ba\~nados metric -- see also Appendix \ref{RobertsAppendix}. 

\begin{figure}
\centering{
\includegraphics[scale=0.45]{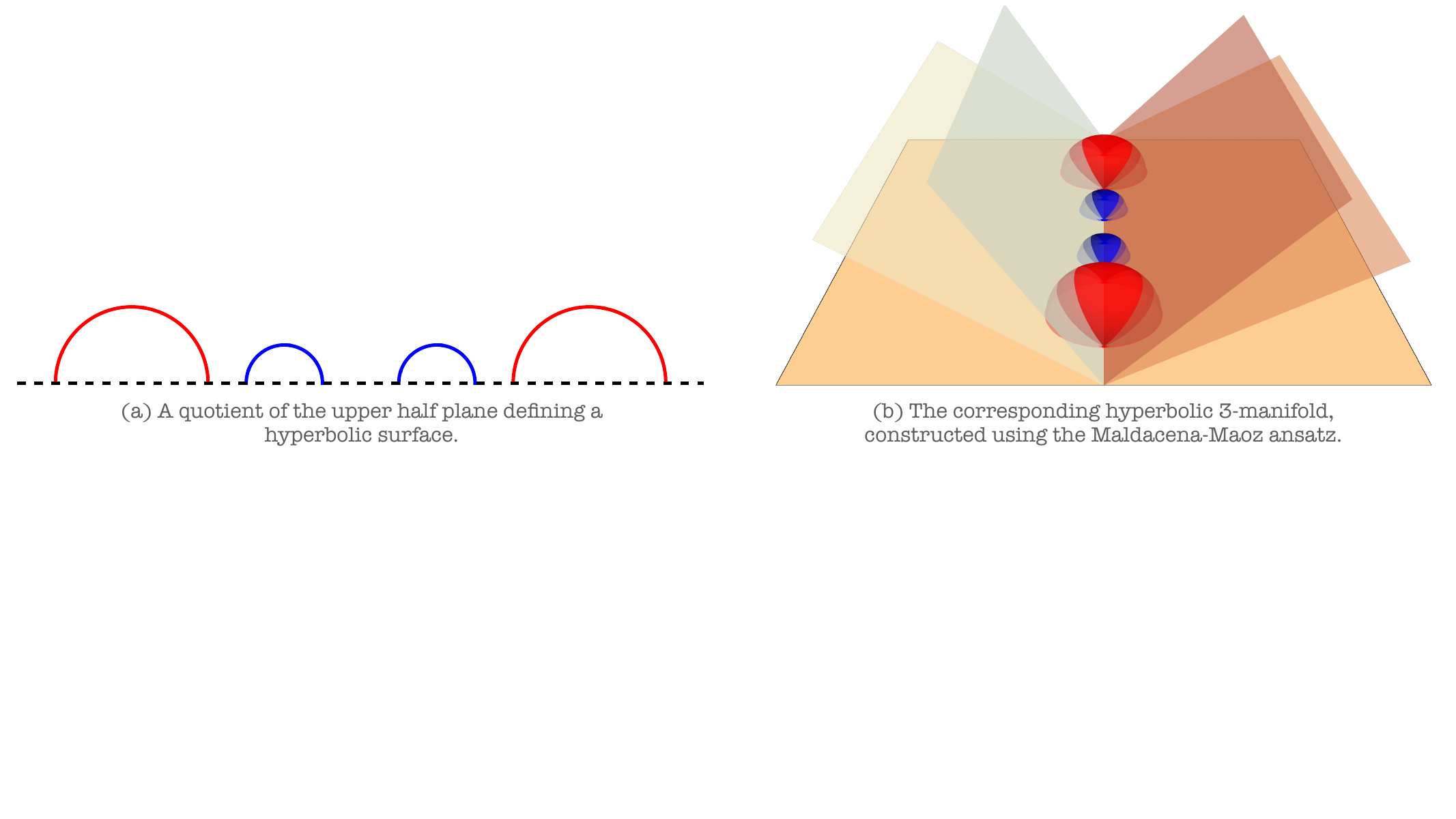}
}
\vspace{-4.0cm}
\caption{The hyperbolic genus-two handlebody as a Maldacena-Maoz wormhole.}
\label{MMCoords}
\end{figure}

Next, we describe features of the Liouville field for the two- and three-point functions in detail.

\subsection*{The Black Hole Two-Point Function}

The Liouville field for the two-point function with black hole insertions at $0$ and $\infty$ follows straightforwardly by substituting $f_{\tt BH}$ and $\bar{f}_{\tt BH}$ from eq.~\eqref{barn9} into eq.~\eqref{eq:liouvilleFromQuotient}. This yields 
\beq
e^{\varphi_{2pt}} = \frac{R_h^2}{w \bar{w}\, \cos^2\! \big(R_h \ln(\sqrt{w \bar{w}})\big)}
\label{Liouville2pt}
\eeq
where $R_h=\sqrt{M-1}$ and recall that $M\ge 1$ for the black hole operators.

Note that $e^{\varphi_{2pt}}$ diverges at  
concentric rings where the denominator vanishes, \ie $w\bar{w}=e^{n\pi/R_h}$ where $n$ is an integer. Scaling $w\to e^{\pi/R_h}w$ leaves the metric \eqref{Kentucky} invariant. Hence book pages are equally well described by allowing $w,\bar{w}$ to run over any one of the anuli between consecutive divergences. 
Further, following the discussion of the AdS$_3$ metric \eqref{H3BookPages}, the divergences correspond to locations where the book pages reach the asymptotic boundary. Hence the book pages reach infinity on two periodically identified intervals, see figure \ref{MMCoords} (b).

At the boundary, we find from \eqref{lio_to_FG} that $w=z$
so we can borrow our discussion around \eqref{radially_to_door} 
to infer what the singularities of $\varphi_{2pt}(w,\bar{w})$ represent in the dome-and-door construction. 
In fact, it follows from that discussion that as move from one operator insertion 
at $w=0$ to the other operator insertion at $w=\infty$, 
we should cross the base of a door infinitely many times, thus the singularities represent precisely the base of the door.

\begin{figure}
\centering{
\includegraphics[scale=0.45]{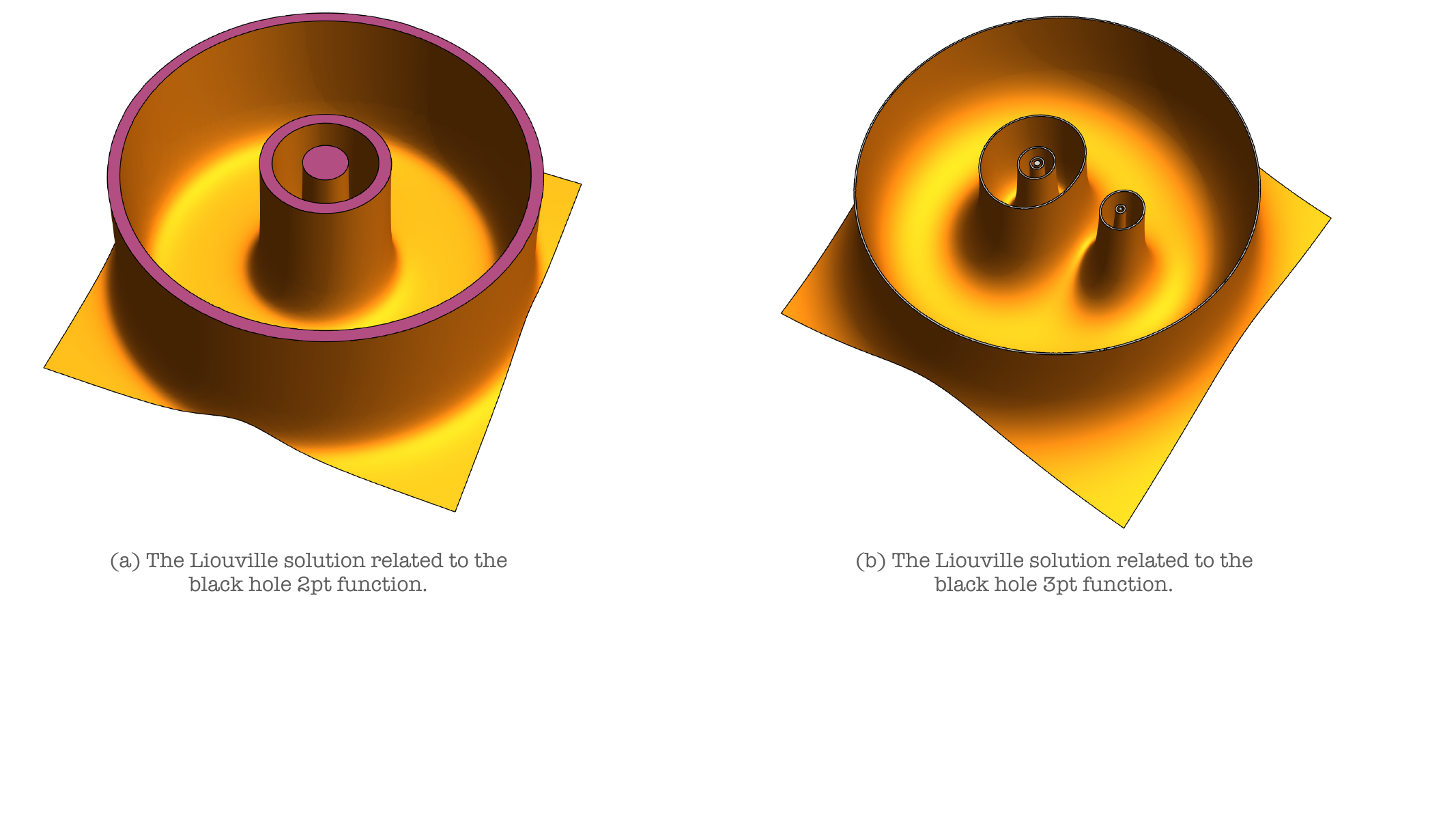}
}
\vspace{-2.5cm}
\caption{$e^{\varphi(z,\bar{z})}$ for the Liouville solutions relevant to  \textbf{(a)} the black hole two-point function and \textbf{(b)} the black hole three-point function.  Around each operator insertion there is an infinite sequence of singularities along closed contours (shown in purple).  These singularities correspond to the location of the doors in the full three-dimensional geometries.  Between each of these singularities is a closed geodesic, corresponding to the black hole horizon.  In the three-point function, there is a single special region which is bounded by three singularity contours rather than only two.  This special chamber also contains three closed geodesics rather than just one.}
\label{liouvilleSols}
\end{figure}

\subsection*{The Liouville Solution with Three Operator Insertions}

For the black hole three-point function with insertion points at $0$, $1$ and $\infty$, 
a solution to the relevant Schwarzian equation \eqref{Roberts_schwartz} is 
the Schwarz triangle function given in eq.~\reef{arc12}.
We can simply plug this Schwarzian solution into eq.~(\ref{eq:liouvilleFromQuotient}) to read off the Liouville solution. 

The result was described in \cite{Hadasz:2003he}.
It has three sets of concentric singularities, each of which accumulates towards one of the operator insertions. The annuli between the singular rings correspond to book pages describing the leg regions, discussed in section \ref{Three}. These book pages are analogous to those found above for the two-point function, and each has two independent boundary regions, corresponding to the singularities in the Liouville field defining the boundaries of the annulus. As with the two-point function solution above, there is a 
scaling symmetry which ensures that all of these annuli describe the same book-page geometry.

There is also a special region that is bounded by three singular contours, one from each of the three families accumulating toward the operator insertions -- see figure \ref{liouvilleSols}(b).
Of course, this special region describes the book page for the central chamber appearing in the dome-and-door construction.

\section{The Action} \label{Action}

We now proceed to evaluate the action for the three-point function geometry discussed in the previous section.

The gravitational action is given by
\beq\label{eq:action}
I = -\frac{1}{16 \pi \GN}\int_\mathcal{M} d\rho \ d^2z \sqrt{g} (R - \Lambda) + \frac{1}{8 \pi \GN} \int_{\partial\mathcal{M}} d^2z \sqrt{h} K + \frac{1}{8 \pi \GN} \int_{\partial\mathcal{M}} d^2z \sqrt{h}
\eeq
where the last term is the counterterm on the asymptotic boundary. Recall we set $L_\mt{AdS}=1$ and hence $\Lambda = -2$. Then, using Einstein's equations, we have $R = -6$.
Furthermore, the extrinsic curvature of the asymptotic cutoff surface is given by $K = -2 + O(\epsilon^3)$.
Therefore, combining the bulk term with the asymptotic boundary contributions in the above action, we find
\beq
I = \frac{1}{4 \pi \GN}\int_\mathcal{M} d\rho \ d^2z \sqrt{g} - \frac{1}{8 \pi \GN} \int_{\partial\mathcal{M}} d^2z \sqrt{h} = \frac{1}{4 \pi \GN} \left(V - \frac{1}{2} A\right)
\label{newaction0}
\eeq
where $V$ and $A$ are the bulk volume and asymptotic boundary area, respectively.
In addition to the contributions \eqref{newaction0}, the action \eqref{eq:action} includes a Gibbons-Hawking-York (GHY) term on the stretched horizons.
So the total action is simply given by
\beq
I = \frac{1}{4 \pi \GN}\left(V-\frac{1}{2} A\right) + \frac{1}{8 \pi \GN} \int_{\text{horizon}} \sqrt{h}K\,.
\label{newaction1}
\eeq

We will need to regulate the volume of the wormhole geometry in order to obtain a finite value for the action.  The correct cutoff is determined by first introducing a ``na\"ive'' cutoff at surfaces of constant $\rho = \pm \ln \left(2 / \epsilon\right)$, and then shifting the cutoff to its correct location. Since we are hoping to obtain a result consistent with a three-point function in a CFT$_2$ in flat space, we would like the metric on the cutoff surface to have the form
\beq
ds_{\text{physical}}^2 = \frac{dz d\bar{z}}{\epsilon^2} + O(\epsilon^0).
\eeq
Note that the na\"ive surface of constant $\rho$ is not the physical cutoff we are interested in.  For instance, it is negatively curved, since the metric on the surface is
\beq
ds_{\text{na\"ive}}^2 = \frac{\left(1 + \frac{1}{4}\epsilon^2 \right)^2}{\epsilon^2} e^{\varphi(z, \bar{z})} dz d\bar{z}.
\eeq
The correct boundary metric can be obtained by shifting the cutoff by an amount proportional to $\varphi(z,\bar{z})$,
\beq
\rho_* = \pm \ln\left(\frac{2}{\epsilon}\right) \mp \frac{\varphi(z,\bar{z})}{2}
\label{arc13}
\eeq

This cutoff cannot be used everywhere, because as explained above the corresponding classical Liouville solution $\varphi(z,\bar{z})$ is singular.  This corresponds directly to the fact that the surface of constant $\rho$ fails to serve as a cutoff at all in cases involving black hole operators, since in such a case all of the constant $\rho$ surfaces reach all the way to the conformal boundary. Hence, we will need to modify the regularization scheme in the region near where these singularities occur.

In \cite{Hadasz:2003kp,Hadasz:2003he}, a prescription for calculating the Liouville action in the presence of three hyperbolic singularities was introduced.
That prescription involves splitting the $z$ plane into a central region, where the Liouville solution is regular, and several ``leg'' regions, where the Liouville solution has singularities.
In the leg regions, the Liouville solution is then replaced by a carefully chosen solution to the \emph{Laplace} equation.
Inspired by this calculation, we will divide our three-dimensional geometry into a central region and several legs.
In the central region, we will find the physical cutoff using the same method that works for the cases involving only conical defects -- see section \ref{Defects} below.  
In each of the legs, we will need to find a new regularization procedure with different ``na\"ive'' cutoffs and shifts to obtain the full physical cutoff. 

\begin{figure}
    \centering
\includegraphics[scale=0.45]{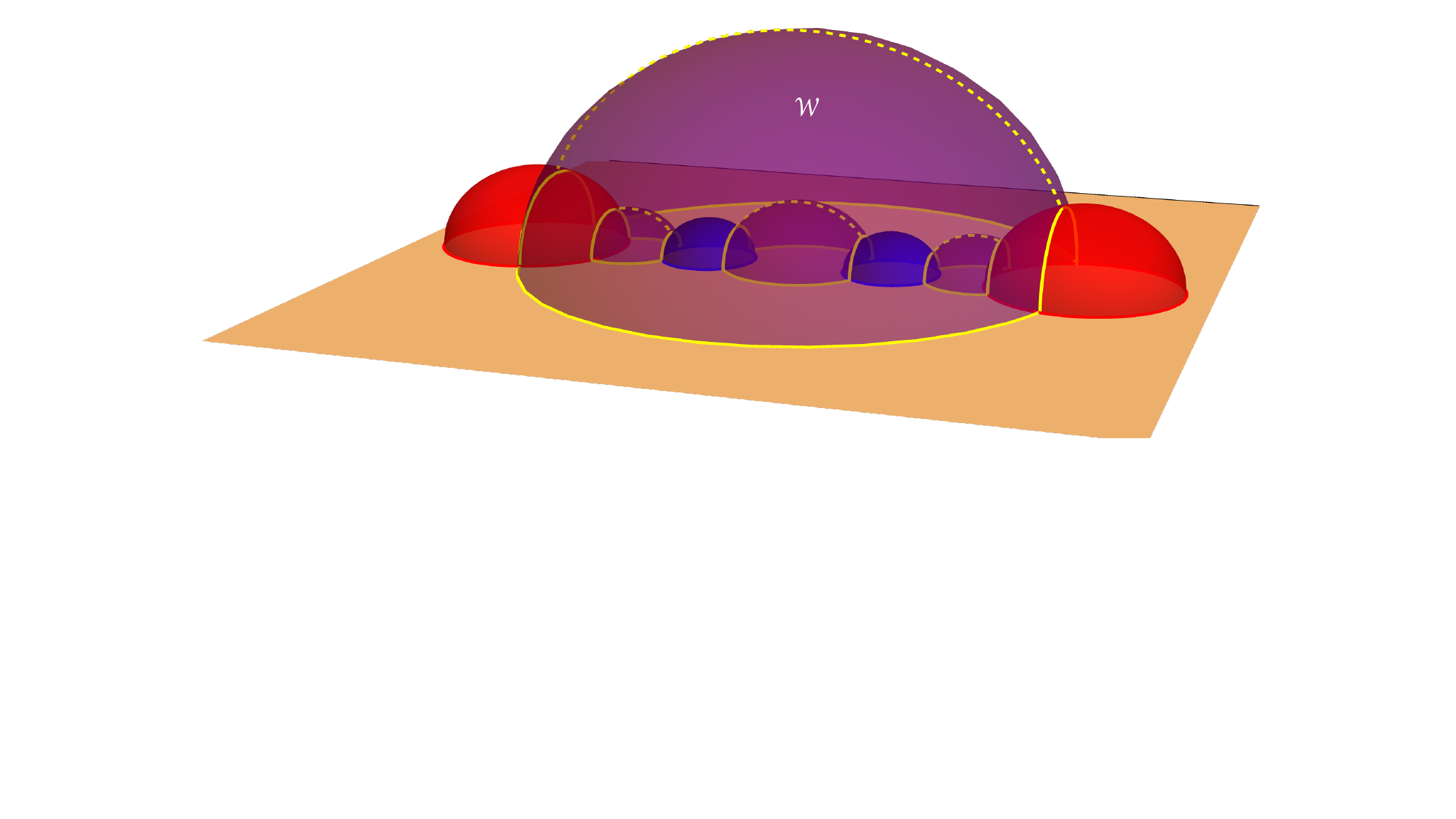}
    \vspace{-4.5cm}
    \caption{The central region $\mathcal{W}$ is defined by first considering the region in $z,\bar{z}$ between the three horizons, shown as dashed yellow lines, on the vertical $\rho=0$ plane. The extension of this region for all $\rho$ is shown here in purple. Passing out of this region through any of the purple walls allows one to approach one of the doors.}
    \label{fig:Wregion}
\end{figure}

To describe the regions more precisely, recall the structure of the metric $e^{\varphi}dz d\bar{z}$ -- see figure \ref{liouvilleSols} (b).
This metric is singular at concentric rings surrounding each operator. 
Between each pair of singularities, there is a closed geodesic corresponding to the horizon of the black hole created by the operator.
In the three point function case, there are three series of concentric rings where the metric is singular (one for each operator).
The three series of concentric singularities are joined by a special ``central chamber'', which is the unique domain in $z,\bar{z}$ that is bounded by three singularities rather than only two.
In this central chamber, there are three closed geodesics (one for each operator), instead of just one.
This corresponds directly to the central chamber in the three-dimensional geometry discussed previously, where one can see all three of the horizons (and all three doors).
The central region $\mathcal{W}$ is the region in the ($z,\bz$)-plane that lies between these three closed geodesics, \ie the three horizons.

Recall from section \ref{stop} that if we continue to Lorentzian signature with $\rho = i t$, we get a spacetime with three asymptotically AdS boundaries, \eg see \cite{Aminneborg:1997pz, Brill:1995jv, Skenderis:2009ju,Balasubramanian:2014hda}.
An observer near any of the three asymptotic boundaries sees a geometry identical to the exterior of a BTZ black hole.
All three of these black holes share a common interior, and 
the central region $\mathcal{W}$ on the $t=0$ slice is exactly this shared interior.

Returning to Euclidean signature, note that removing the region $\mathcal{W}$ from the ($z,\bz$)-plane for all $\rho$ leaves behind six disconnected pieces of the geometry. These are the six ``leg'' regions, $\mathcal{L}_i$.
It is inside these leg regions where the $\rho=\rho_*$ cutoff given in eq.~\reef{arc13} fails, since each leg contains one of the concentric series of singularities.
So we will need a new prescription to regulate the action in the leg regions.

\begin{figure}
\centering{
\includegraphics[scale=0.45]{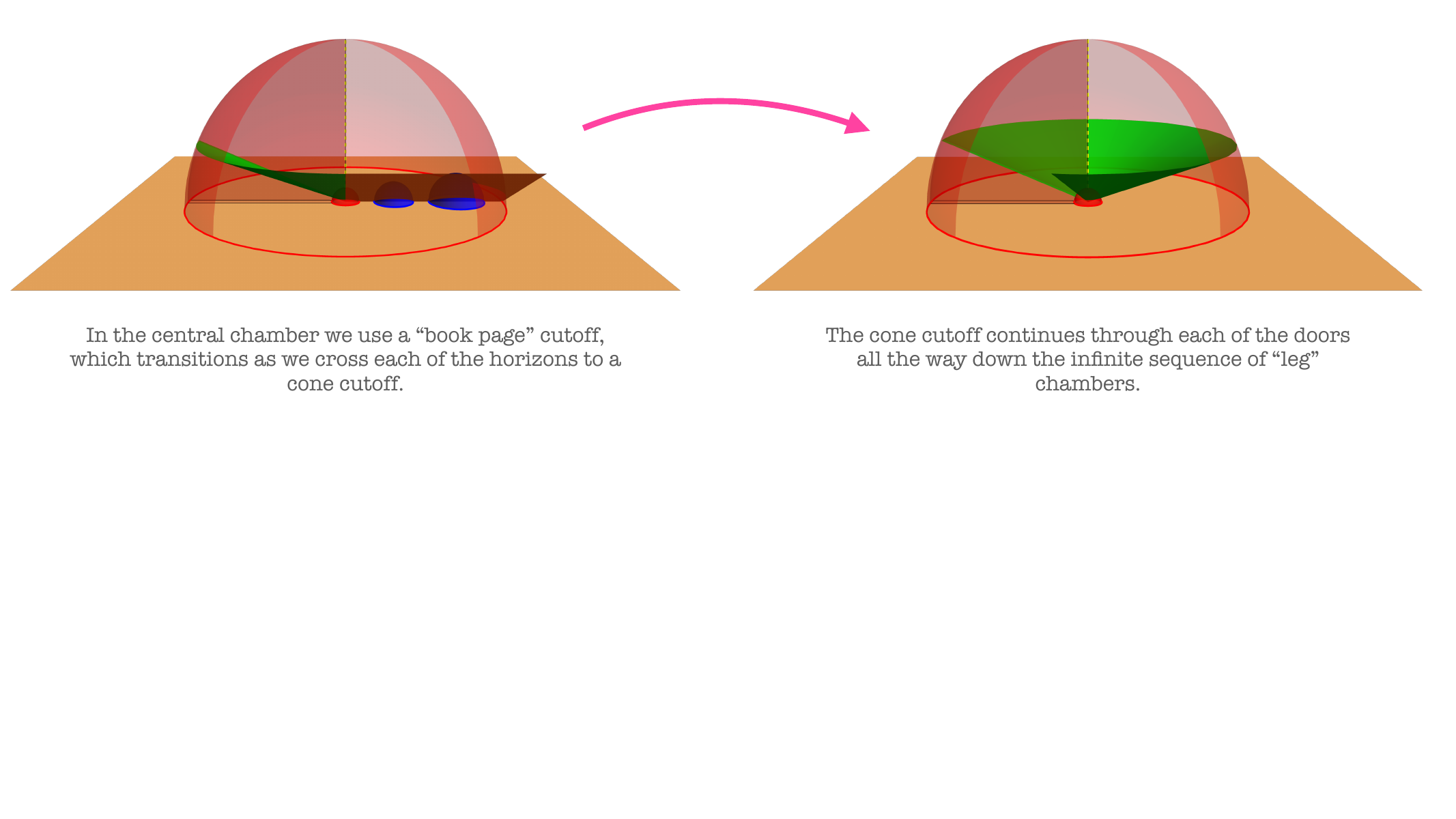}
}
\vspace{-5.0cm}
\caption{We use different ``na\"ive'' cutoffs in the central region and the leg regions.  Each of these are then shifted to obtain a cutoff with the correct metric.  These shifts are controlled by solutions to the Liouville and Laplace equations in the central and leg regions, respectively.}
\label{3ptCutoffs}
\end{figure}

The na\"ive cutoff which we will use in each of the leg regions is given by a surface which is isometric to a cone in the \{$Y$, $Z$, $\bar{Z}$\} coordinates.  That is, by performing an isometry, we can choose the na\"ive cutoffs for each of the legs to be cones in coordinates in which the metric takes the form (\ref{emptyMetric}) -- see figure \ref{3ptCutoffs}.  So, they are most easily described by introducing another system of coordinates on Euclidean AdS$_3$,
\beq
\begin{split}
Y = e^{\tau}\, \text{sech}\,\tilde{\rho}\,, \\
Z = e^{\tau+ i \theta}\, \tanh\tilde{\rho}\,, \\
\bar{Z} = e^{\tau - i \theta}\, \tanh\tilde{\rho}\,.
\end{split}
\label{arc14}
\eeq
In these coordinates, the metric takes the form
\beq
ds^2 = d \tilde{\rho}^2 + \cosh^2\!\tilde{\rho}\, d\tau^2 + \sinh^2\!\tilde{\rho}\,d\theta^2\,.
\label{eq:legMetric}
\eeq
In these coordinates, the na\"ive cutoff for each of the leg regions is then a surface of constant $\tilde{\rho}$, which is a cone in the Poincar\'e coordinates (\ref{emptyMetric}). The coordinates $\tau$ and $\theta$ will be related to the physical coordinates in the CFT with
\beq
\tau = \frac{1}{2}\left(\tilde{f}_i(z) +\bar{\tilde{f}}_i(\bar{z}) \right)\,, \qquad
\theta = \frac{1}{2 i} \left(\tilde{f}_i(z) - \bar{\tilde{f}}_i(\bar{z}) \right)\,,
\label{ace8}
\eeq
where $i$ labels the six leg regions.

Similarly to what we saw in the central region, the na\"ive cutoff has the wrong metric.  So we will use a cutoff related to this na\"ive one by a similar shift
\beq
\tilde{\rho}_* = \ln\left(\frac{2}{\epsilon}\right) - \frac{\tilde{\varphi}_i(z,\bar{z})}{2}
\label{arc15}
\eeq
where $\tilde{\varphi}_i(z,\bar{z})$ are each solutions to the {\it Laplace} equation, given by
\beq
\tilde{\varphi}_i(z,\bar{z}) = \ln\left(\partial \tilde{f}_i\,\bar{\partial}\bar{\tilde{f}}_i\right)\,.
\eeq
With these choices, the metric on the cutoff surface is the desired flat metric
\beq
ds^2_{\text{cutoff}} = \frac{dz d\bar{z}}{\epsilon^2} + O(\epsilon^0)\,.
\eeq
Finally, in order for the full cutoff to be smooth, we need to impose gluing conditions at each of the boundaries $\Gamma_i$ between the central and leg regions:
\beq
\begin{split}
\varphi(z,\bar{z})|_{\Gamma_i} = \tilde{\varphi}_i(z,\bar{z})|_{\Gamma_i}\,, \\
\partial \varphi(z,\bar{z})|_{\Gamma_i} = \partial\tilde{\varphi}_i(z,\bar{z})|_{\Gamma_i}\,, \\
\bar{\partial} \varphi(z,\bar{z})|_{\Gamma_i} = \bar{\partial}\tilde{\varphi}_i(z,\bar{z})|_{\Gamma_i}\,.
\end{split}
\label{ace9}
\eeq

To summarize, we have split the geometry into several regions -- see figure \ref{fig:Wregion}.
In each region, we have a natural way to describe the physical cutoff as the shift of a ``na\"ive'' cutoff -- see figure \ref{3ptCutoffs}.
In the central region, corresponding to the $\mathcal{W}$ in the ($z,\bar{z}$)-plane, we will use the cutoff $\rho = \rho_*$ in eq.~\reef{arc13}.
This cutoff has a shift by the Liouville field of a constant $\rho$ ``bookpage''.
In the leg regions $\mathcal{L}_i$, we  use the cutoff $\tilde{\rho}=\tilde{\rho}_*$ given in eq.~\reef{arc15}.
There, the shift is made by a solution to the Laplace equation of the constant $\tilde{\rho}$ cone.
Although we are describing these cutoffs in different coordinates, the gluing conditions on the Liouville and Laplace solutions guarantee that we are actually describing a single continuous cutoff surface.

\subsection*{Central Region Action} 

We start by considering the ``central'' region corresponding to $\mathcal{W}$ defined above.
Using the expression for the book page metric (\ref{Kentucky})-(\ref{eq:liouvilleFromQuotient}) and integrating over $\rho$, we find that the volume of the central region is
\beq
\begin{split}
V = \int d^2 z \left(\frac{1}{\epsilon^2} + \ln\left(2/\epsilon\right)e^{\varphi} - \bar{\partial}(\varphi \partial\varphi) + \partial\varphi \bar{\partial}\varphi \right)
\end{split}
\eeq
where we used the equations of motion, $\partial\bar{\partial}\varphi = \frac{1}{2}e^{\varphi}$. 
Plugging in the induced metric on the asymptotic cutoff surfaces, $\rho= \pm\rho_*$ gives
\beq
A = \int d^2 z \left(\frac{2}{\epsilon^2} + e^{\varphi} + \partial \varphi \bar{\partial} \varphi \right).
\eeq
Combining these expressions as in eq.~\reef{newaction0}, the action for the central region is given by
\beq
\begin{split}
I_{\mathcal{W}} = \frac{1}{8 \pi \GN}\int_{\mathcal{W}} d^2 z
\left(\partial\varphi \bar{\partial}\varphi + e^{\varphi}\right) - \frac{i}{8 \pi \GN} \oint_{\partial\mathcal{W}} \!\!\!dz\ \varphi \partial\varphi - \frac{1}{2 G}\left(1 - \ln(2/\epsilon)\right)\,.
\end{split}
\label{one1}
\eeq
In the last term, we used the fact that the area of the wormhole region is\footnote{This can be seen by remembering the realization of $\mathcal{W}$ as a quotient of the hyperbolic upper half-plane. 
Then we see that $\mathcal{W}$ is a hyperbolic octagon with all right angles (with some edges identified).
Then the result (\ref{WArea}) can be derived using the general formula for the area of a hyperbolic polygon: $A = (n-2) \pi - \sum_i \a_i$
where $n$ is the number of sides and $\a_i$ are the interior angles of the polygon.}
\beq\label{WArea}
\int_{\mathcal{W}}d^2 z \ e^{\varphi(z,\bar{z})} = 2\pi\,.
\eeq
We see that, up to the final term, which will be cancelled by counterterms, the action of the central region is the Liouville action for $\varphi(z,\bar{z})$ in the corresponding subregion of the $(z,\bar{z})$-plane, $\mathcal{W}$.  This is very similar to what happens in the conical defect case, as discussed in \cite{Chandra:2022bqq}. 

\subsection*{Leg Region Action}

Now consider the contribution to the action of one of the ``leg'' regions, where the metric takes the form (\ref{eq:legMetric}).
Integrating over $\tilde{\rho}$, we find that the total volume is
\beq
\begin{split}
V = \int d^2 z \left(\frac{1}{2 \epsilon^2} - \frac{1}{4} e^{\tilde{\varphi}_i}\right)
\end{split}
\eeq
For the boundary contribution, we need the area of the surface $\tilde{\rho} = \tilde{\rho}_*$ 
as defined in eq.~\reef{arc15}.
Then $A$ is given by
\beq
A = \int d^2 z \left(\frac{1}{\epsilon^2} + \frac{1}{2}\partial \tilde{\varphi}_i \bar{\partial} \tilde{\varphi}_i \right)\,.
\label{two2}
\eeq
Combining the bulk and boundary terms together as in eq.~(\ref{newaction0}), we find
\beq
I_{\mathcal{L}_i} = \frac{1}{16 \pi \GN}\int_{\mathcal{L}_i} d^2z\left(\partial \tilde{\varphi}_i \bar{\partial} \tilde{\varphi}_i - e^{\tilde{\varphi}_i}\right) - \frac{i}{8 \pi \GN} \oint_{\partial\mathcal{L}_i}\!\!\! dz\ \tilde{\varphi}_i \partial \tilde{\varphi}_i\,.
\label{arc16}
\eeq
where we used the equations of motion $\partial \bar{\partial}\tilde{\varphi}=0$, and integrated by parts to extract a boundary term that will combine nicely with the one which we found in the central region.

We see that the action in the each leg region is given by the Liouville action evaluated on a solution to the \emph{Laplace} equation $\tilde{\varphi}_i(z,\bar{z})$ and with the opposite sign for the Liouville cosmological constant (the minus sign in front of $e^{\tilde{\varphi}_i}$ in the first integral above).
It may seem strange at first that we are evaluating the Liouville action of a solution to the Laplace equation, but in fact it matches perfectly with the proposal in \cite{Hadasz:2003he,Hadasz:2003kp}, except for the wrong-sign cosmological constant.
Having the correct sign for the Liouville cosmological constant is crucial for reproducing the expected dependence of the three point function (squared) on the location of the operator insertions.
Thankfully, eqs.~\reef{one1} and \reef{two2} are not the only contributions, since as we argued in \cite{Paper1}, we must also include a GHY term on the stretched horizons, as we have already indicated in eq.~\reef{newaction1}.
The inclusion of this boundary term makes the geometries we are considering analogous to the fixed area states of \cite{Akers:2018fow, Dong:2018seb} (see also \cite{Dong:2022ilf}), since it allows us to fix the metric on the horizon.
We will see soon that the inclusion of the Gibbons-Hawking-York term on the stretched horizons exactly flips the sign of the Liouville cosmological constant!

\subsection*{Gibbons-Hawking-York Term on the Horizons}

We need to include a GHY terms on each of the stretched horizons.  The stretched horizons are simply the surfaces with $\tilde{\rho} = \epsilon$.  Computing the induced metric on these surfaces, we find that the volume form is
\beq
\sqrt{h} \ d\tau d\theta = \epsilon \ e^{\tilde{\varphi}_i} dz d\bar{z} + O(\epsilon^3),
\eeq
while the extrinsic curvature is given by simply $K = \frac{1}{\epsilon} + O(\epsilon)$.

\begin{figure}
    \centering
    \includegraphics[scale=0.48]{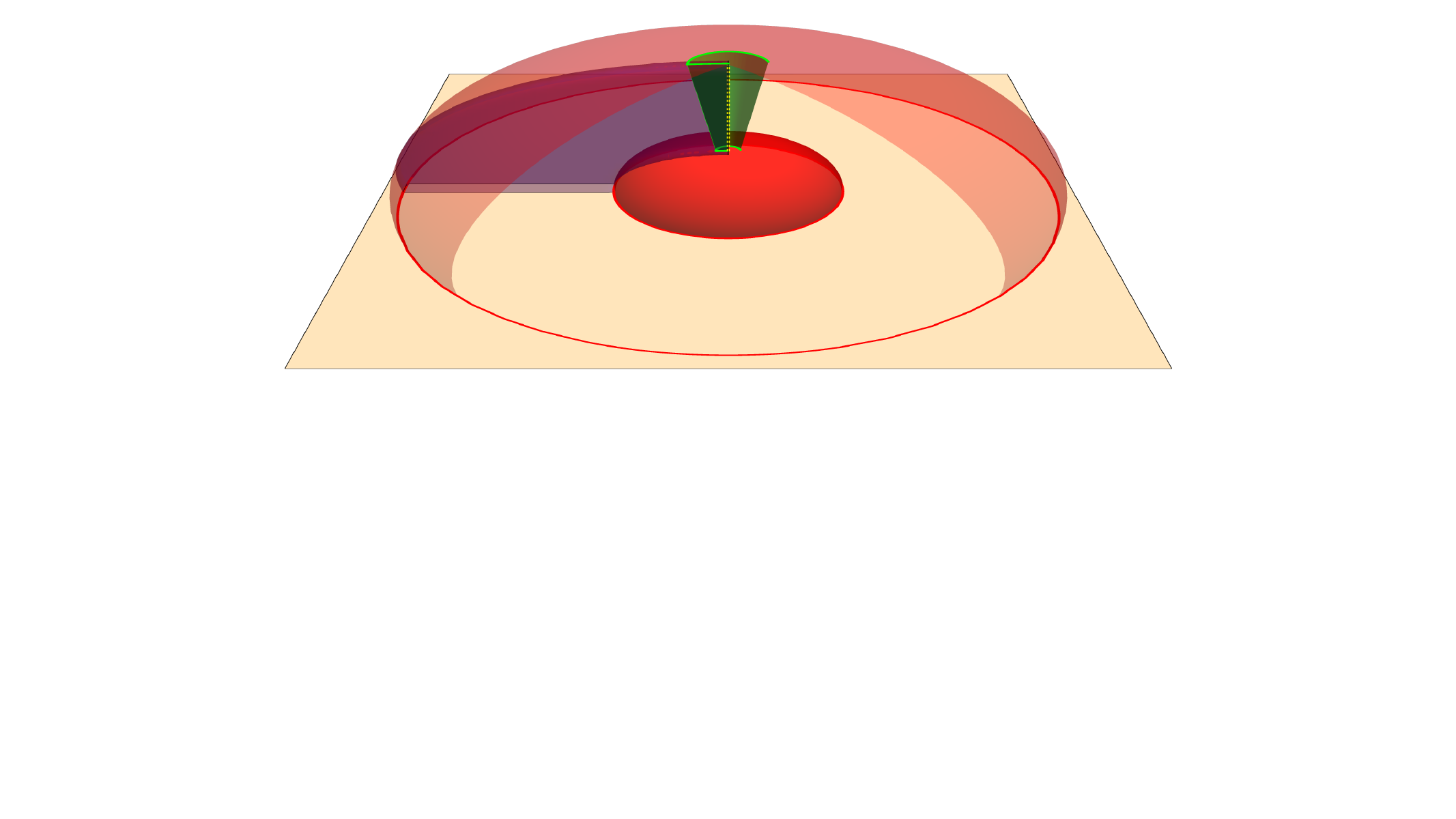}
    \vspace{-6cm}
    \caption{The stretched horizon (green cone)  crashes into the true horizon (yellow, dashed line) before it goes through the door and enters the central chamber.}
    \label{crashingHorizon}
\end{figure}

The stretched horizons extend along each of the six legs.
We also need to decide what to do with the stretched horizon in the central region $\mathcal{W}$.
We will make what we consider a ``minimal'' choice.
That is, we will have the stretched horizon crash into the true horizon just before it enters the central chamber -- see figure \ref{crashingHorizon}.\footnote{This choice may seem unusual and our primary justification will be the agreement of our results with the semiclassical Liouville three-point function. However, see further discussion in section \ref{Defects}. }

With this choice, the GHY terms at the horizons give a contribution only in the leg regions, and we find
\beq
I_{\text{GHY,hor}} = \frac{1}{8 \pi \GN}\sum_i \int\sqrt{h} K = \frac{1}{8 \pi \GN}\sum_i \int_{\mathcal{L}_i} d^2 z \ e^{\tilde{\varphi}_i}\,.
\label{three3}
\eeq
Combining this to the previous leg contributions in eq.~\reef{arc16}, we see that the sign of the Liouville potential in the leg region is flipped.  

Combining eqs.~\reef{one1}, \reef{two2} and \reef{three3}, we have
\begin{flalign}
    \begin{aligned}
I &= I_{\mathcal{W}} + \sum_i\( I_{\mathcal{L}_i} + I_{\text{GHY,hor}}\) \\
&= \frac{1}{8 \pi \GN}\int_{\mathcal{W}} d^2 z
\left(\partial\varphi \bar{\partial}\varphi + e^{\varphi}\right)  + \frac{1}{16 \pi \GN} \sum_{i}\int_{\mathcal{L}_i} d^2z\left(\partial \tilde{\varphi}_i \bar{\partial} \tilde{\varphi}_i + e^{\tilde{\varphi}_i}\right)   \\
&\qquad- \frac{i}{8 \pi \GN} \sum_i  \oint_{|z-z_i|=\epsilon}\!\!\!\! dz \ \varphi_i \partial\varphi_i  - \frac{i}{8 \pi \GN} \oint_{|z|=1/\epsilon} \!\!\!dz \ \varphi \partial \varphi - \frac{1}{2 \GN}\left(1 - \ln(2/\epsilon)\right).
    \end{aligned}&&&
    \label{arc17}
\end{flalign}
Notice that the boundary terms from \reef{one1}  have nicely cancelled with the ones in \reef{arc16}, since they lie along the common boundary of $\mathcal{W}$ and the $\mathcal{L}_i$'s.
This is the importance of the gluing conditions \reef{ace9}.

As mentioned before, a procedure for calculating the action of the Liouville solution with three hyperbolic singularities was introduced in \cite{Hadasz:2003he,Hadasz:2003kp}.
Their result has the correct dependence on the locations of the operator insertions, and the position independent term in their result matches with the semiclassical limit of the DOZZ formula.
Their calculation involves the introduction of solutions to the Laplace equation, which can be identified with our $\tilde{\varphi}_i(z,\bar{z})$.  They define the regularized Liouville action as
\beq
\begin{split}
&I_L = \frac{1}{2 \pi}\int_{\mathcal{W}} d^2 z
\left(\partial\varphi \bar{\partial}\varphi + e^{\varphi}\right)  + \frac{1}{2 \pi} \sum_{i=1}^3\left(\int_{\mathcal{L}_i} d^2z\left(\partial \tilde{\varphi}_i \bar{\partial} \tilde{\varphi}_i + e^{\tilde{\varphi}_i}\right)  \right) \\
&\qquad- \frac{i}{\pi} \oint_{|z|=1/\epsilon}\!\!\! dz \ \varphi \partial \varphi - \frac{i}{\pi} \sum_{i=1}^3  \oint_{|z-z_i|=\epsilon}\!\!\!\! dz \ \tilde{\varphi}_i \partial\tilde{\varphi}_i  - \left(4 + \sum_{i=1}^3(1 - R_i^2)\right)\ln(\epsilon)\,.
\end{split}
\eeq
Comparing with our result \reef{arc17}, we see
\beq
I = \frac{c}{6}I_L - \frac{c}{3}(1-\ln(2)) + \frac{c}{6}\sum_{i=1}^3(1 - R_i^2) \ln(\epsilon)
\eeq
where we used the relation $c = \frac{3}{2 \GN}$.

The remaining dependence on the cutoff is canceled when we add the required counterterms, which can be found by examining the two point function -- see Appendix \ref{2ptApp}.
Hence in the end, we have
\beq\label{actionFinal}
I = \frac{c}{6}I_L - \frac{c}{3}(1-\ln(2))
\eeq
As discussed in \cite{Chandra:2022bqq}, the second term arises from the dependence of the DOZZ formula on the Liouville cosmological constant $\mu$ when one scales it as $\mu = \frac{1}{4 \pi b^2}$ in the semiclassical limit, $b \rightarrow 0$.  Hence we find
\beq
e^{-I} \approx |G_L(z_1,z_2,z_3)|^2
\label{arc18}
\eeq
where $G_L(z_1,z_2,z_3)$ is the semiclassical Liouville three-point function with a specific normalization related by analytic continuation to the one used in \cite{Chandra:2022bqq}, and we have checked the agreement at the classical level. Of course, this result \reef{arc18} with the square of the three-point function is aligned with our discussion in section \ref{WormH}, where the appearance of the wormhole suggested that the geometric calculation yields a contribution to the variance 
of the three-point function.

Now we turn to discussion of the relationship between the black hole three-point function geometry and geometries involving defect operators.
We will then return to the discussion of the Lorentzian cap, and consider its contribution to the action in more detail.

\section{Geometric Transitions. Doors and Defects.}\label{Defects}

\begin{figure}
\centering{
\includegraphics[scale=1.5]{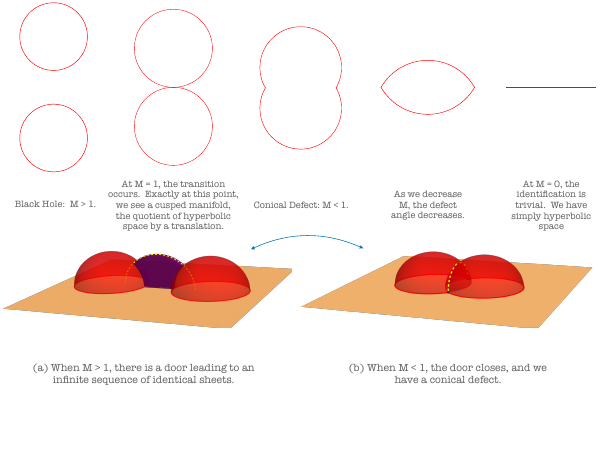}
}
\vspace{-1.75cm}
\caption{{\bf Top:} Boundary picture. As we decrease the mass of the operator insertions, a transition occurs in which the black hole becomes a conical defect.  The defect angle for a given mass, $M$, is $2 \pi \left(1- \sqrt{1-M}\right)$.  Eventually, as we continue to decrease $M$, the defect disappears. {\bf Bottom:} Full picture. When the transition between black hole and conical defect is made, the door closes.  In the left figure, the yellow dashed line is the black hole horizon, around which the geometry branches infinitely many times, as represented by the purple door.  In the right figure, the yellow dashed line represents the conical defect, which appears at the intersection of the domes defining the identification.}
\label{transition}
\end{figure}

Since we are discussing gravity in three dimensions, there is a gap in energy between the vacuum and the first black hole state.
For an operator to be dual to a black hole, we need that its mass to be $M > 1$.
So far we have been considering only such black hole operators.
However, it is straightforward and illuminating to extend the constructions above to the case where some of the operators have mass $0<M<1$, which are dual to conical defects in the bulk.

\begin{figure}
\centering{
\includegraphics[scale=0.45]{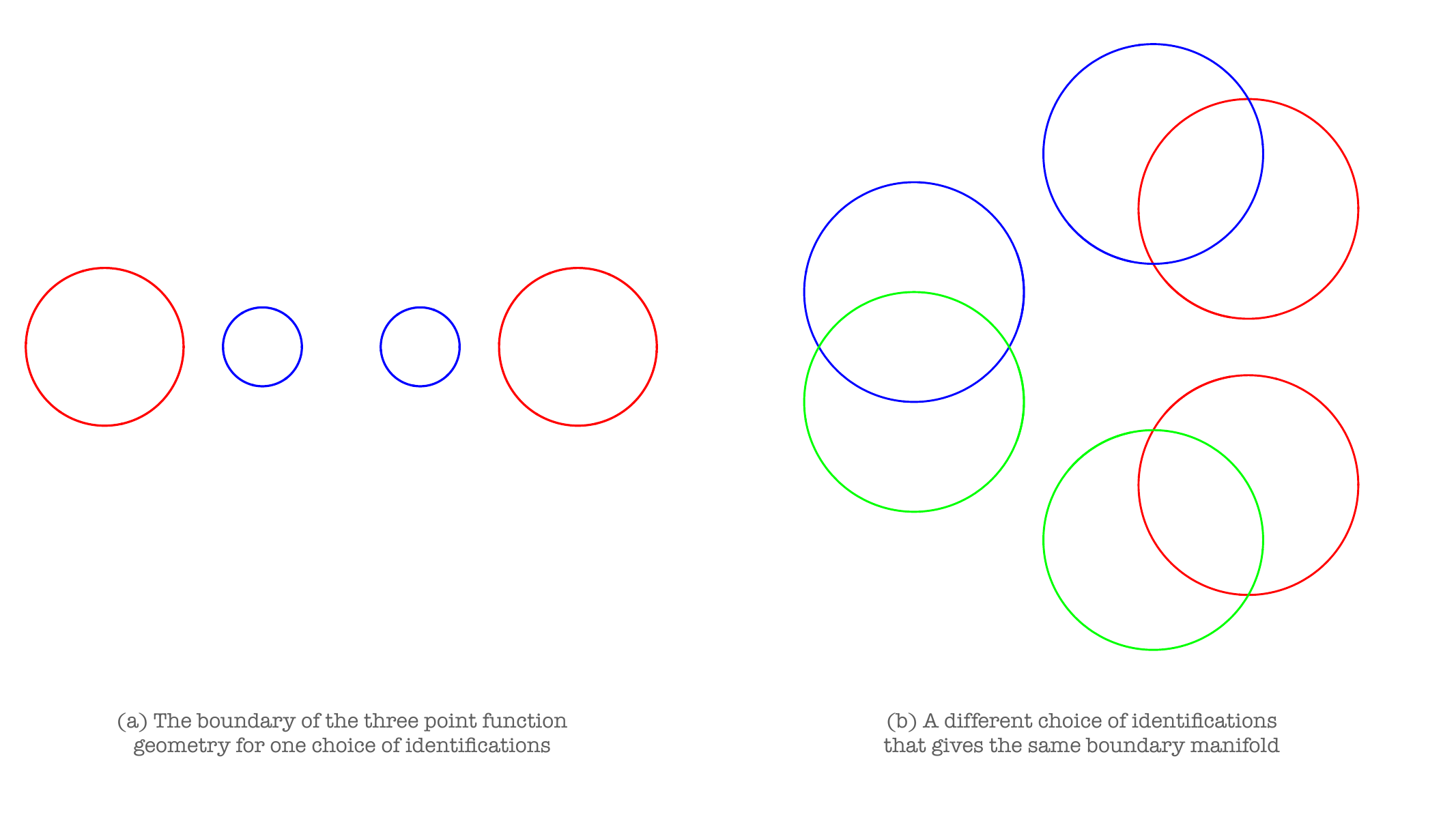}
}
\vspace{-0.5cm}
\caption{Two different choices of branch cuts for the boundary of the three-point function geometry with three black hole operator insertions. Circles with the same color are identified, creating the same genus-two surface in both cases.}
\label{3ptIdents2d}
\end{figure}

Consider the two-point function geometry.
As we decrease the mass of the operators, the domes used to represent the corresponding identification of hyperbolic space approach each other.
As we pass below $M=1$, the identification used to construct the geometry goes from being a hyperbolic isometry to an elliptic one. In terms of the dome pictures, the transition occurs as the domes collide --  see figure \ref{transition}.
The resulting intersection of the two domes is the trajectory of a conical defect.

In terms of the Schwarzian solutions, the transition is also apparent.
Recall for the black hole case (with $M>1$), we had
\beq
f_\texttt{BH}(z) = z^{i\sqrt{M-1}}\,.
\eeq
The fact that the image of the $z$ plane under this function covers the ($Z,\bZ$)-plane many times is what necessitated the introduction of the door.
On the other hand, with $M<1$, the exponent is \textit{real}, \ie
\beq
f_\texttt{defect}(z)=z^{\sqrt{1-M}}\,.
\eeq
The image covers only a portion of the $Z$ plane once, and the door has disappeared.

\begin{figure}
\centering{
\includegraphics[scale=0.45]{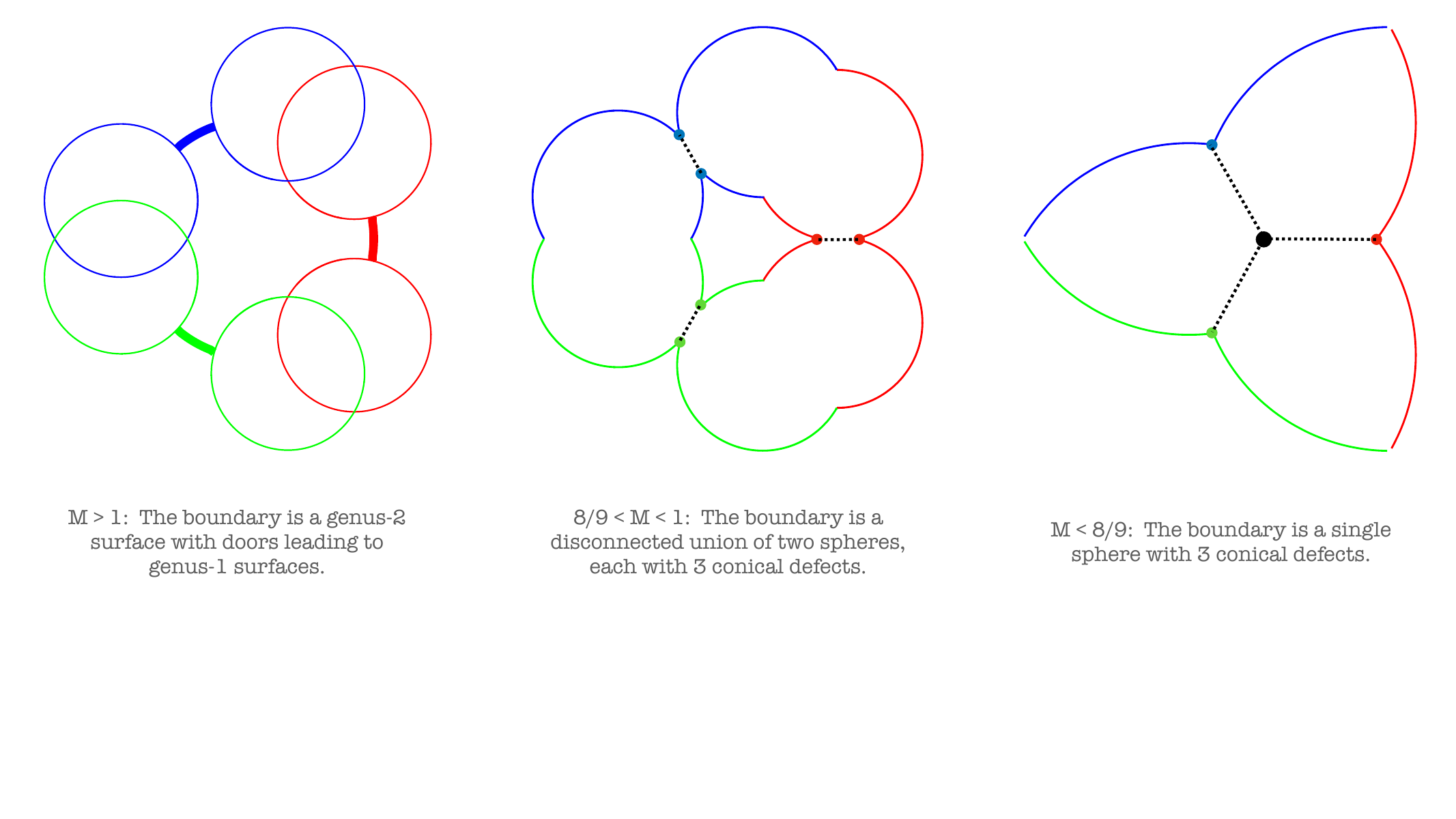}
}
\vspace{-2.7cm}
\caption{The three-point function geometry for three equal-mass operators, seen from above. 
When M > 1, the identifications create a genus-two surface, which has three doors (thick lines).  These divide the asymptotic boundary into two regions (\ie we have a wormhole). 
As we decrease the mass so that $M<1$, the black holes are replaced by defects (dashed lines).  The doors collapse and close, but the boundary is still divided into two pieces.
If we decrease the mass even more, so that $M< 8/9$, the wormhole itself disappears, and the defects join at a vertex (the black dot in the center of the right figure) in the bulk instead. Although they appear flat here, the dashed lines are actually semicircular arcs shooting through the bulk.}
\label{fig:3ptTransition2d}
\end{figure}

This story continues naturally to the three-point function case.
Here, it is easier to understand the transitions by considering a different choice of fundamental domain -- see figure \ref{3ptIdents2d}.
The two panels correspond to different choices for the branch cuts of the Schwarzian solution $f_{\texttt{3pt}}$.
With the choice of identifications shown in figure \ref{3ptIdents2d}(b), we can follow what happens as we decrease the masses of all three operators -- see figure \ref{fig:3ptTransition2d}.
We see that first, the circles corresponding to a given identification collide, the doors disappear and they are replaced by conical defects at the intersections of the corresponding domes -- see figure \ref{3ptIdents2d}(b).
These conical defects shoot into the bulk, and travel to another point on ``the'' asymptotic boundary.
However, just as in the black hole case, the boundary is actually divided into two disconnected components, in this case without the need for doors.
Thus, we see that we have conical defects flying from one side of a two-sided Euclidean wormhole to the other.
This geometry was studied in \cite{Chandra:2022bqq}, and
its gravitational action was found to reproduce the square of the expected universal HHH three-point function.

As we decrease the masses even more, a second transition occurs when 
\beq
\sum_i \sqrt{1-M_i} = 1\,.
\eeq
At this point, the area of the throat of the wormhole $\mathcal{W}$ goes to zero.
The wormhole closes up, and instead of travelling from one asymptotic boundary to another, the defects meet at a point in the bulk -- see figure \ref{fig:3ptTransition2d}(c).
Here we have the single-sided solution studied in \cite{Chang:2016ftb}, where they also found the universal result for the boundary structure constant.

It is also straightforward to consider mixed cases of defect-BH-BH or defect-defect-BH geometries.
In these cases, the calculation of the action proceeds with a mixture of what we have just done for the three black hole case and what was done in \cite{Chandra:2022bqq} for three defects.
We introduce a stretched horizon with GHY term for each black hole operator, and we can choose whether to consider a two-sided geometry or a single-sided one with a Lorentzian cap.
The result again has the universal form of the HHH three-point function.

\begin{figure}[t]
\centering{
\includegraphics[scale=0.57]{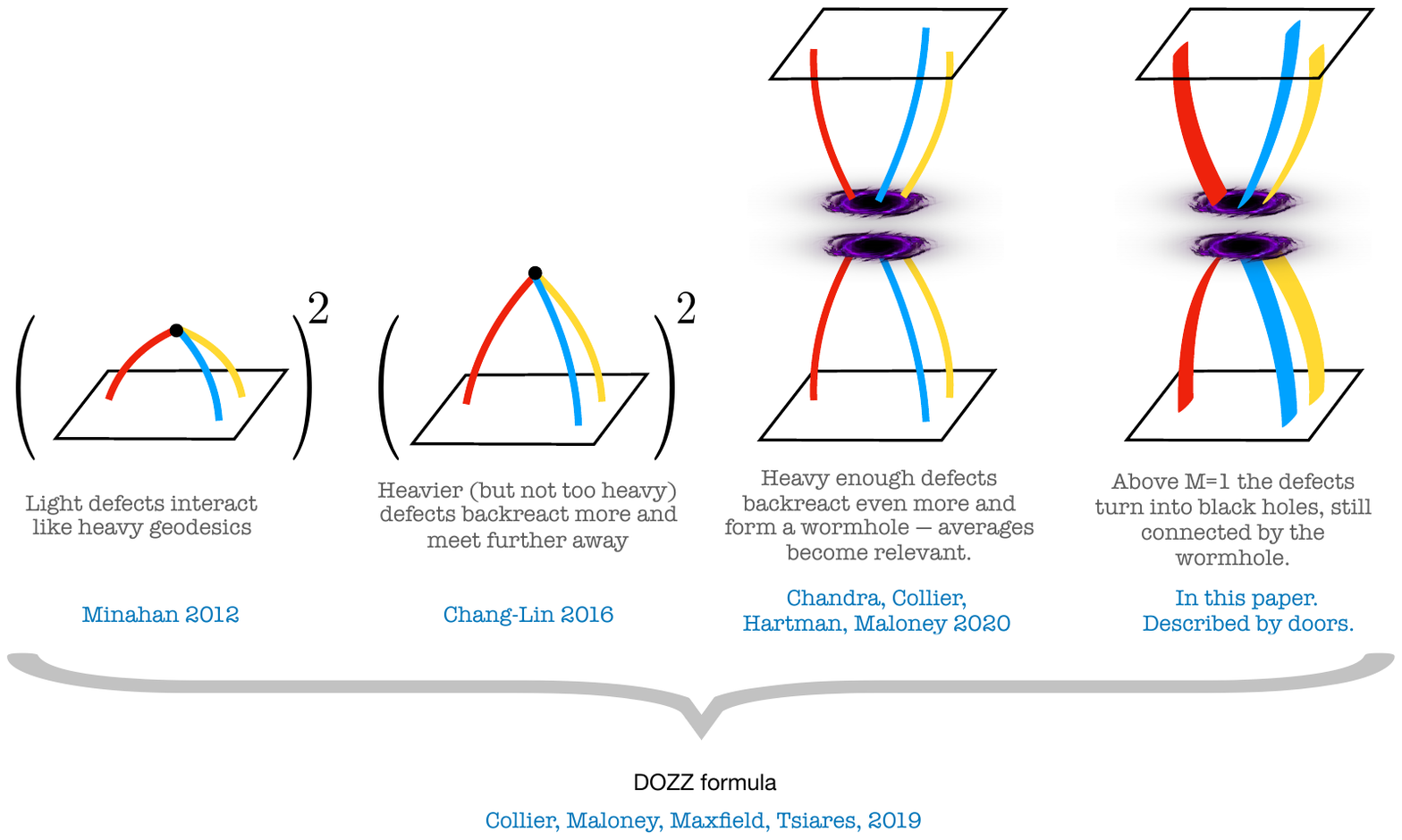}}
\vspace{-3cm}
\caption{Heavy operator correlators in $AdS_3/CFT_2$ are universally given by the DOZZ formula \cite{Collier:2019weq}. Nonetheless, their bulk geometrical description varies quite a bit. Very light defects can be well described by geodesics meeting at an optimal point which minimizes their lengths; as the defects become heavier they interact further from the boundary; these solutions were studied by Chang and Lin in \cite{Chang:2016ftb}. Eventually they cross a threshold where they would interact at infinity; beyond this point the solution forms a wormhole and becomes double sided, becoming the solution studied by Chandra, Collier, Hartman and Maloney \cite{Chandra:2022bqq}. For even heavier masses, they become black holes and we get a geometry with black holes traversing a wormhole studied in this paper. The dome picture described herein unifies all these geometries and their interpolations.   
} \label{DOZZ}
\end{figure}

The complete story is summarized in figure \ref{DOZZ}.

\subsection*{Lorentzian Caps and Single-sided Geometries}

Now we would like to consider the proposal in section \ref{stop} where continuing the mouth of the wormhole to a Lorentzian cap turns our two-sided Euclidean solution into a single-sided solution.
Since the cap is Lorentzian, the corresponding action $I_{\texttt{cap}}$ will contribute to the three-point function as a phase
\beq
G_3(z_1,z_2,z_3) = e^{-I_{\texttt{Euclidean}} + i I_{\texttt{cap}}}
\eeq
The key result is that the contribution of the Lorentzian geometry to the action, $I_{\texttt{cap}}$ is a sum of separate contributions from each of the three operators.

\begin{figure}
    \centering{
    \includegraphics[scale=0.48]{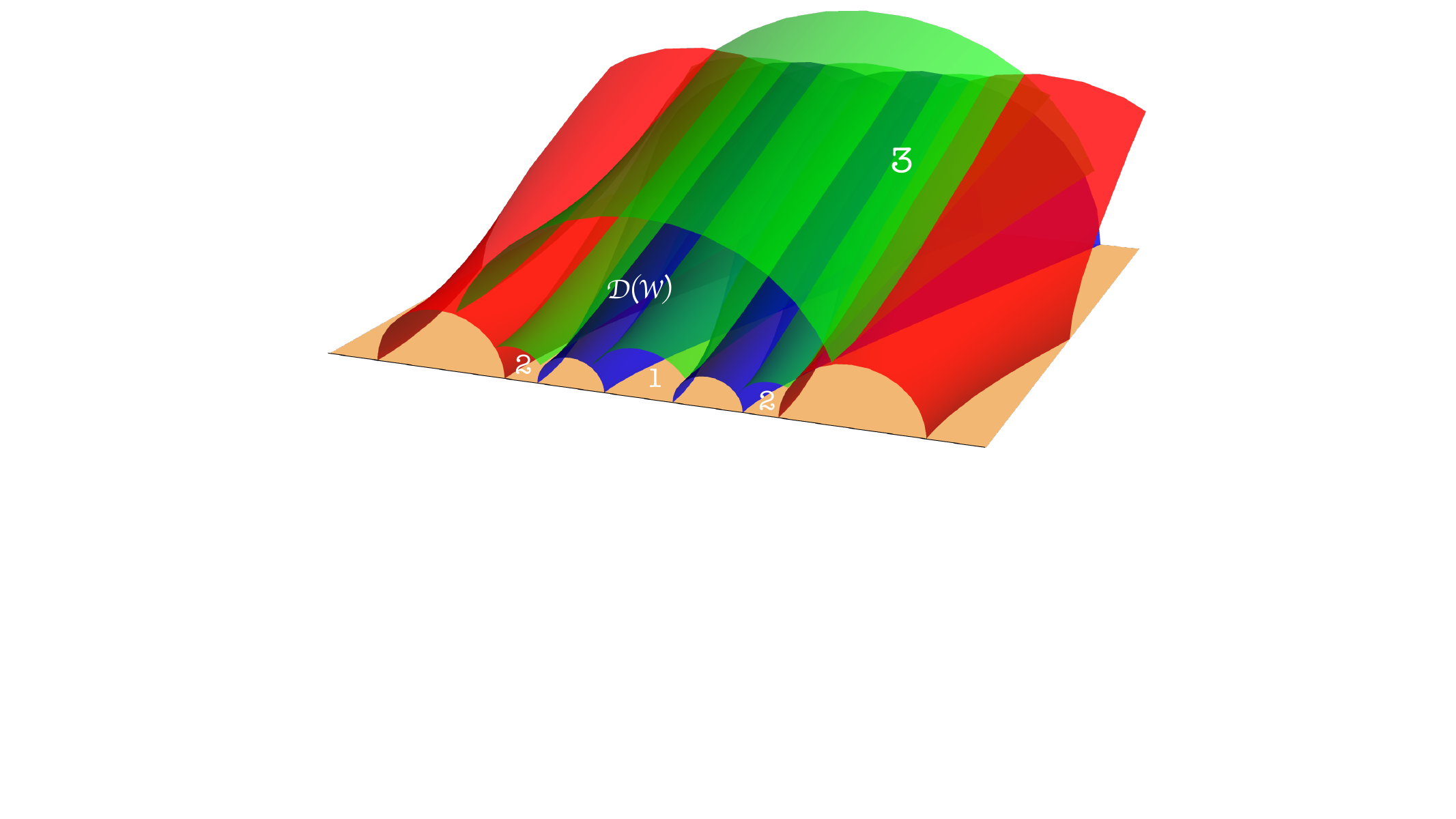}
    \vspace{-5cm}
    \caption{Here we show the Lorentzian cap for the 3pt geometry. The domain of dependence of $\mathcal{W}$ is the region between the green surfaces, while the rest of the upper half space lies outside of it. We see that removing the domain of dependence splits the spacetime into three disconnected regions.
    We associate the three disconnected regions each to one of the operators, and absorb the corresponding phase factor into the definition of the operator.} \label{fig:domainOfW}
    }
\end{figure}

The Lorentzian cap has a metric obtained by performing a Wick rotation as in eq.~\ref{new-bookpage} on the wormhole metric \reef{dino1},
\beq
-dt^2 +\cos^2\!t\  e^{\varphi(z,\bar{z})} \ dz d\bar{z}\,.
\eeq
We glue this Lorentzian geometry with $t>0$ onto the $\mathcal{W}$ region (\ie the region between to three horizons) at $\rho=0$.

The area of the throat of the wormhole (\ie the region $\mathcal{W}$) at time $t=0$ is
\beq
\int_{\mathcal{W}} d^2z \ e^{\varphi(z,\bar{z})} = 4 \pi \sum_{i=1}^3 \text{Re}(\eta_i) - 4\pi\,.
\eeq
where $\eta_i = \frac{1}{2} - \frac{1}{2} \sqrt{1-M_i}$. Note that if $\eta_i$ corresponds to a conical defect (with $M_i<1$), then $\text{Re}(\eta_i) = \eta_i$, while if $\eta_i$ corresponds to a black hole (with $M_i>1$), then $\text{Re}(\eta_i) = \frac{1}{2}$.
One contribution to the volume of the Lorentzian cap is given by the volume of the region $\mathcal{W}$ for $0<t<\pi/2$.
This contribution is given by
\beq
V_{\mathcal{W}} = \int_0^{\frac{\pi}{2}} dt \ \cos^2(t) \int_{\mathcal{W}} d^2z \ e^{\varphi(z,\bar{z})} = \pi^2 \sum_{i=1}^3 \(\text{Re}(\eta_i) - \frac13\)\,.
\eeq
The crucial property of this result is that it is a sum of three separate contributions that each depend only on the properties of a single operator.

In the conical defect case, this is the full cap.
However, if the geometry contains a black hole, the region $\mathcal{W}$ for $0<t<\pi/2$ does not cover the full cap.
It contains only the domain of dependence of $\{z \in \mathcal{W}, t=0\}$. However, the geometry at the surface at $t=\pi/2$ is smooth, \ie it is only a coordinate singularity, and so we may choose to include further portions of the Lorentzian geometry beyond this surface. For example, 
we choose to include the entire interior region of the black hole(s). Again it is crucial to note that the additional geometry will be comprised of a disconnected component for each black hole operator -- see figure \ref{fig:domainOfW}.
Since these regions lie outside of the domain of dependence of $\mathcal{W}$, their geometry (\eg whether they contain shockwaves or any other disturbances) are not determined by the boundary condition we have supplied.
If we simply analytically continue the given solution, the geometry of these regions is identical to the Einstein-Rosen bridge in a two-sided BTZ geometry with the same mass as the corresponding operator.
We will take the interpretation that the details of what appears outside the domain of dependence of $\mathcal{W}$ is part of the definition of the operator associated to that region.
So in sum, the action of the Lorentzian cap is
\beq
I_{\text{cap}} = \frac{1}{4 \pi \GN} V_{\mathcal{W}}+\sum_{i=\text{BH}} I_i = \frac{\pi}{4 \GN}\sum_i \(\text{Re}(\eta_i)-\frac13\) + \sum_{i=\text{BH}} I_i \,.
\eeq
Hence we see that it is the sum of three contributions that each depend only on one of the three operators.
Thus, we can absorb the corresponding phase generated by the action of the Lorentzian geometry into the definition of the operators, and we will arrive at a real result for the three-point function. 

Hence we arrive at the total action for the single-sided geometry,
\beq
I_{\text{single-sided}} = \frac{1}{2} I_{\text{wormhole}}+ i \sum_i f(i)\,.
\eeq
Again, it is a nontrivial result that the imaginary term is given by the sum shown above since this allows us to absorb the corresponding
phases into the definition of the operators. After eliminating these phases, we recover the three-point function from the single-sided bulk geometry, \ie
\beq
e^{-I} \approx G_L(z_1,z_2,z_3)\,,
\label{arctic18}
\eeq
where $G_L(z_1,z_2,z_3)$ is the semiclassical Liouville three-point correlator  in \cite{Chandra:2022bqq}.

\section{Discussion} \label{discussion}

In this paper we discussed three-dimensional asymptotically AdS$_3$ geometries that 
are sourced by the insertion of boundary operators whose scaling dimensions 
is heavy as the central charge of the holographic CFT$_2$. 
The presence of any such operators deforms the AdS geometry by inducing a non vanishing expectation value 
for the holographic stress tensor, close to the boundary. 
This is true perturbatively in general dimensions, 
but in three-dimensions there is an exact solution, due to Ba\~nados \cite{Banados:1998gg},
that describes such deformation.
However, this metric does
not describe the full bulk spacetime.
When only two black hole operators are inserted, 
we showed that the full geometry is simply
an infinite covering of the Euclidean BTZ black hole \cite{Paper1}, but
when three or more operators are inserted, 
we found that the completion of the Ba\~nados metric into the bulk 
is a wormhole geometry involving multiple asymptotic boundaries.
To understand this rather non trivial fact we rephrased the construction of the bulk geometry 
as a quotient of AdS$_3$ realized by domes and doors. The dome construction 
is a well know characterization of hyperbolic geometries with an asymptotically AdS$_3$ metric, 
and more familiar from the study of black hole thermodynamics, see \eg \cite{Maxfield:2016mwh},
but the addition of the doors is new as far as we can tell.  

As in the description of a Euclidean two-point function 
geometry in section \ref{Two}, \ie as empty AdS$_3$ with identifications,
the doors are needed to describe the insertion of boundary operators. In particular, 
approaching a black hole operator insertions 
means circling around a dome and thus going through the door infinitely many times.
When a third operator is inserted, as we discussed in section \ref{Three}, the presence of the doors splits the 
AdS$_3$ boundary into multiple asymptotic boundaries and a wormhole appears. 
An interesting alternative to the Euclidean 
wormhole\footnote{We thank Juan Maldacena for bringing this possibility to our attention.} 
is to cut the geometry at the wormhole mouth and glue there a Lorentian cap. In this way 
the bulk on-shell action can be understood to compute a holographic three-point function, up to phases. 
We expect that geometries with $n$-black holes operator insertions are again Euclidean wormholes, 
since it is straightforward to extend our dome-and-door construction to examine the case with more than three insertions.
It would be interesting to understand how these geometries, constructed here using domes and doors, would be described in the language of the recently proposed ``Virasoro TQFT'' of \cite{Collier:2023fwi}.

\begin{figure}[t]
\centering{
\includegraphics[scale=0.57]{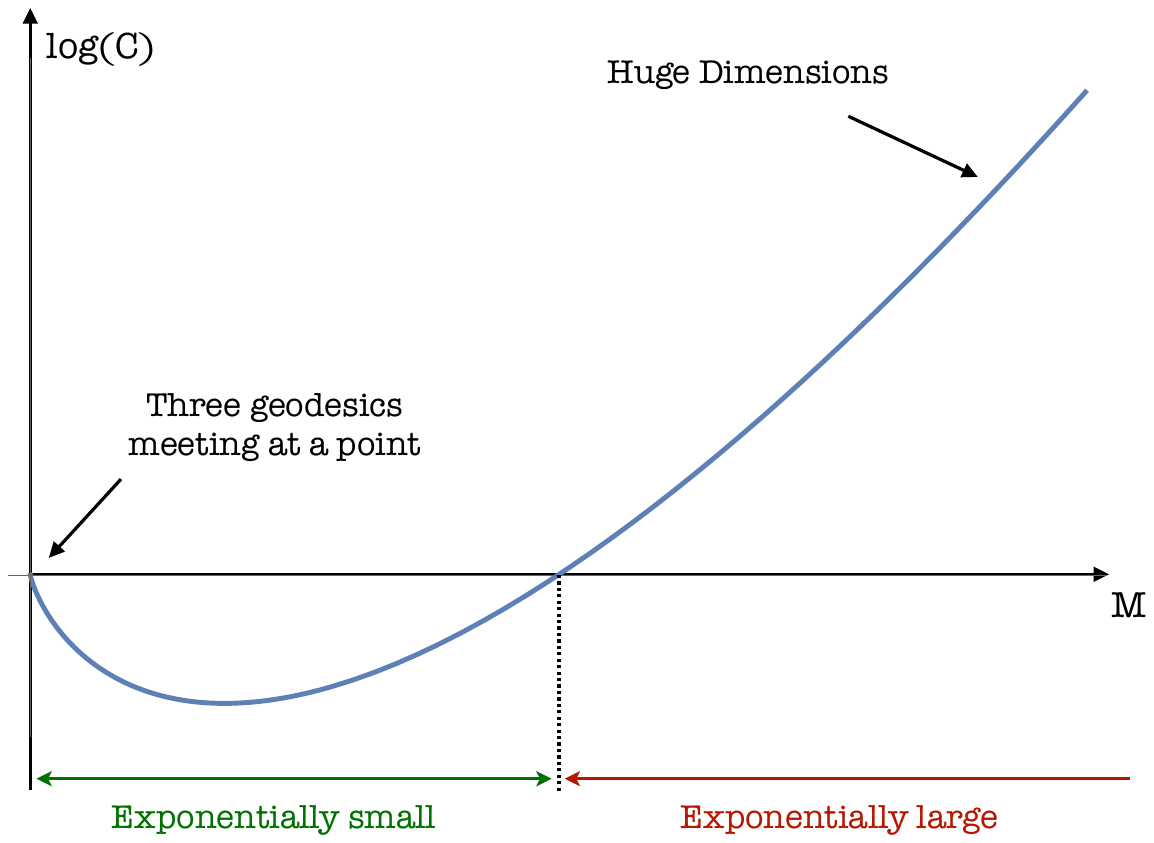}}
\vspace{-3cm}
\caption{As we go from small to large masses (\ie conformal dimensions), $\log C$ changes sign. The structure constant is thus exponentially small for ``small'' masses and exponentially large for ``large'' masses. 
} \label{cartoonDOZZ}
\end{figure}

For discussion, it is instructive to consider some simplifying regimes. 
For example, things simplify when the operators are either very light or very heavy.
In the first case, three black holes should behave as pointlike probes moving along geodesics meeting at an optimal point in the bulk. 
Then the correlator should be approximated by \textit{minus} 
the mass times length of these geodesics so that \cite{Minahan:2012fh}
\beqa
\log C_{123} &\simeq& \sum_{i\neq j\neq k} \frac{\Delta_i}{2} \log\left(\frac{(\Delta_i - \Delta_j + \Delta_k)(\Delta_i +\Delta_j-\Delta_k)(\Delta_i+\Delta_j+\Delta_k)}{4\Delta_i^2(\Delta_j+\Delta_k-\Delta_i)}\right) \nn \\
&=& - \frac{3}{2}\, \Delta \log(4/3) < 0  \qquad \text{for equal masses}
\eeqa
For very heavy operators, we also expect things to simplify and indeed 
we observe that the DOZZ formula as function of insertions points, and 
at very large conformal dimensions $\Delta_i$, can be nicely captured by the simple integral
\beq
\log |G_L(z_1,z_2,z_3)|^2 \simeq \frac{c}{3 \pi}\int d^2 z \sqrt{L\bar L} \la{integralLL}
\eeq
This integral can be  evaluated straightforwardly and matched with the well known asymptotics of DOZZ; it would be great to understand how to derive it directly from a simple gravitational argument (something we partially achieve below where we mention the analytic continutation to negative masses).
For  $\Delta_i=\Delta$, the above \eqref{integralLL} gives
\cite{Cardy:2017qhl}
\beq
\log C_\texttt{BH BH BH} \simeq  \frac{3\Delta}{2} \log(27/16) > 0 \la{CBHBHBH}
\eeq
Note that $ C_\texttt{BH BH BH}$
is exponentially large, while $C_{123}$ for geodesics 
is exponentially small. The interpolation is drawn schematically in figure \ref{cartoonDOZZ}. 
Perhaps  there is a simple CFT explanation for such a drastic change 
in the scaling of the three-point coupling.\footnote{Note that if we consider a correlation function of very large operators 
in a gauge theory, we often get exponentially large results precisely 
indicating that the number of contractions (\ie the entropy) is huge. 
For example, see the discussion of $C_{123}$ for three operators dual 
to LLM geometries in \cite{Paper1}. There the result is exponentially 
large for any R-charges~$J_i \sim N^2$.} 

Collecting the results from sections \ref{Two}, \ref{Three} and \ref{Defects} 
we provided a uniform geometric picture 
for interpolating between these extreme regimes. In particular 
we discussed the geometry as we vary from the
regime in which the geometry sourced by 
the operator insertions matches the result for the geodesic computation, 
and it grows into finite defect operators, until 
the regime where the geometry describes black hole operators, 
for which we have found a wormhole.

A more general consideration that our three-dimensional explorations are strongly demanding is the following: What will happen in 
higher dimensions when we try to extend a three-point geometry 
beyond the Fefferman-Graham patch, all the way into the bulk? 
and what geometric picture underlies the interpolation between geodesics and huge operator insertions?
Will we find a wormhole, 
or will we rather find a single horizon with the topology obtained 
by fattening three geodesics meeting at a point into a three-legged sphere? 
Perhaps neither is true and something even more exotic will arise. 
In \cite{Paper1}, we also advocated the possibility of a conifold-like solution 
with three spacetime bananas.
It would be fascinating to develop new analytical or numerical 
techniques for finding the three-point function geometry in higher dimensions.  

Relatedly, in \cite{Benjamin:2023qsc}, the authors predict 
that the typical structure constant of three very heavy global primary operators, 
with equal dimensions $\Delta_i=\Delta$, is at leading order independent from the spacetime dimensions, and reads
\beq 
\log C_{\texttt{Heavy Heavy Heavy} } \simeq \frac{3\Delta}{2} \log(36/16) \label{CHHH}
\eeq
What is the holographic counterpart of this very unusual universality? 
We do not know of any other example where a physical quantity computed 
by a gravitational action is independent from the spacetime dimensions.

Another observation highlighted in \cite{Benjamin:2023qsc} is that by reconsidering \eqref{CHHH} in two dimensions, it follows that the structure constants for primaries is much larger than (\ref{CBHBHBH}), which is the leading order result for the structure constants of huge Virasoro primaries that we obtain holographically as explained in this paper.
The fact that we are reproducing (\ref{CBHBHBH}) and not (\ref{CHHH}) makes sense since the stress-tensor expectation (\ref{L}) we start with is the expectation of $T$ in the presence of three Virasoro primaries.
It would be very interesting to construct the \textit{typical} stress tensor expectation value (\ref{barn7}) when $O_\Delta$ are global primaries corresponding to descendants of Virasoro primaries with a level comparable to the dimension and redo the analysis of this paper with that effective $L(z)$ as starting point.\footnote{Is the level is $O(1)$ the expectation value will lead to the very same (\ref{L}) as can be checked by a straightforward 2d CFT computation in the semiclassical limit where $c,\Delta$ scale together to infinity.}
Will that reproduce (\ref{CHHH})? Can that be translated to a boundary graviton computation as discussed in \cite{Benjamin:2023qsc}? Will that computation better hint at the higher $d$ generalization for which we only have the preliminary ideas alluded to in \cite{Paper1}?

Perhaps some intuition can be obtained from the following observation. The large dimension prediction of  \cite{Benjamin:2023qsc} can be cast as the statement that the structure constant of three huge operators of dimension $\Delta_i$ should be approximately given by a very simple ratio of factorials  
\beq
C_{\texttt{Heavy$_1$ Heavy$_2$ Heavy$_3$}} \simeq \frac{(\Delta_1+\Delta_2+\Delta_3)!}{2^{\Delta_1}(\Delta_1)! \times 2^{\Delta_2}(\Delta_2)!\times 2^{\Delta_3}(\Delta_3)!} \label{combin}
\eeq
as can be checked by comparing the Stirling approximation of the square of 
this expression to the leading expectation value for three heavy operators 
contained in the second line of~eq.~(7.51) in \cite{Benjamin:2023qsc}. 
Clearly, (\ref{combin}) is begging for a combinatorial interpretation! If 
we interpret the denominator as coming from the normalization of the operators 
and the numerator as coming from the genuine interaction 
the task is to figure out how to generate $(\Delta_1+\Delta_2+\Delta_3)!$. 
Is there any (set of) Feynman diagrams that could produce such simple dependence 
in a large $N$ gauge theory for instance?\footnote{One naive suggestion would 
be to have one big bulk vertex with $\Delta_1+\Delta_2+\Delta_3$ fields 
connected to the three boundary operators, each with $\Delta_i$ fields. 
The number of ways to pick constituents from this bulk vertex to connect 
to the boundary would then easily produce such factorials. This is naive 
because vertices with $n$ indistinguishable legs would naturally come 
with a $1/n!$ prefactor thus cancelling this factorial. } 
Can the topology of such diagrams hint at a dual gravity picture? 
Since the main interaction factor only depends on the sum of all dimensions, 
this naively would suggest that the three operators merge into some 
big object in the bulk, akin to the interaction meeting point of three light geodesics, in contradistinction with the wormhole picture we encountered for Virasoro primaries.\footnote{This is also quite 
different from the combinatoric result arising from the interaction 
of three fully symmetric LLM large operators as computed in \cite{Paper1}. 
There the combinatorics final result took the form of a product of terms 
depending on effective dimensions $\Delta_{ij}=(\Delta_i+\Delta_j-\Delta_k)/2$ 
hinting as a sort of splitting of each external LLM geometry into two 
effective geometries of dimension $\Delta_{ij}$ with pairwise interactions between these effective geometries taking place.} 

Another regime where things could simplify and help our intuition is the Heavy-Heavy-Light correlators 
when the light operator is not neutral, implying that the two heavy operators can not 
be the same by charge conservation \cite{Escobedo:2011xw,Yang:2021kot}. 
Then the light operator is most probably not just a probe geodesics. 
It would interesting to understand the implications of this statement on the gravity side.

Finally, let us add an intriguing observation concerning negative masses as 
a way to explore the physics of huge mass operators, at least holographically.
The idea is to explore this unphysical regime as a  trick to study the analytic continuation of the 
on-shell action for potisive and large masses. We shall see now that this idea allows for 
a shortcut derivation of the asymptotics (\ref{CBHBHBH}). 
The key observation is that when black holes 
have negative mass, most, if not all of the geometry can be reached by staying below the \text{wall} 
(the surface where the Ba\~nados metric has zero determinant). We explain this
in detail in appendix \ref{negativeM}. The simplest scenario to discuss 
is the two-point function with insertions at zero and infinity.
There the $\det(g) = 0$ wall is a cone which when mapped back to global coordinates 
is located at some finite $r> r_h$ when $M$ is 
positive, \eg see discussion around eq.~(3.9) in \cite{Paper1} and 
appendix \ref{negativeM}.
But when $M<0$ the wall actually corresponds to $r=0$, which is the location of the naked singularity. 
Thus, the Ba\~nados patch covers the full geometry in this case! 
A na\"ive estimate of the bulk geometry in this case then reads
\beq
-I = \frac{c}{6 \pi}\int_{\epsilon/x <|z| < x/\epsilon} d^2 z \, \sqrt{L(z)\bar L(\bar z)} = \frac{c |M|}{12} \int_{\epsilon/x}^{x/\epsilon} \frac{dR}{R}  \frac{1}{R} = -2\Delta \log(x/\epsilon)
\eeq
This contribution alone gives
$
e^{-S}=|x/\epsilon|^{c |M| /6}=|x/\epsilon|^{-2 \Delta}
$
which is the expected growth 
when $M<0$. Importantly, it gives precisely the 
analytic continuation of the full $M>0$ result derived 
in \cite{Paper1}. 

For the Ba\~nados banana  with two fixed insertions 
at $z_1$ and $z_2$ -- instead of zero and infinity -- the picture is a bit more subtle, but
the geometry there is still almost all outside the wall. In particular, 
when $M \to -\infty$ the volume of the region inside the wall is subleading, 
as explained in Appendix \ref{negativeM}. Thus we reproduce again the expected result.

The negative mass two-point function discussion immediately leads to the plausible picture that 
perhaps the same will happen for the three-legged geometry: 
As $M_i \to -\infty$, the geometry outside the wall is almost the full geometry.
Then it is straightforward to compute the bulk action from the Ba\~nados patch and 
one obtains that the action on-shell is related to the integral of $\sqrt{L \bar L}$. 
For three points we immediately reproduce \eqref{integralLL} and 
thus \eqref{CBHBHBH} for large negative mass. If we assume no essential 
singularity at infinity, the large positive mass result follows by 
analytic continuation. This is a much simpler derivation than the one from the Liouville 
field in section \ref{Action}! Does a similar shortcut yield the 
asymptotic large mass behaviour of higher point functions as well 
in terms of the integral of the corresponding~$\sqrt{L \bar L}$'s?\footnote{Of course, the result for higher point functions would not be as universal. For a four point function, 
for instance, we have $$L(z)=-\frac{1}{(z-z_1)(z-z_2)(z-z_3)(z-z_4)} \left(\sum_{i}\frac{M_i \prod_{j\neq i}z_{ij}}{4(z-z_i)}+U\right)$$ where the constant $U$ is a theory dependent 
constant which is not a simple function of the dimensions and central charge. } 

Also, is there a counterpart of this picture for large negative mass in higher dimensions? Should it  lead to (\ref{CHHH})?

Maybe not; perhaps for global primaries things are more subtle (not only for large masses but for any masses). Maybe the typical three point function of huge CFT operators corresponds to a gravitational picture of three black holes dressed by a complicated cloud of matter and gravitons as speculated in \cite{Benjamin:2023qsc}. Conversely, according to that scenario -- which we are not necessarily endorsing or finding any sort of evidence for in our explorations -- the holographic dual of the three huge black holes geometry is \textit{not} just a typical three point function of very heavy operators in the dual CFT. What would it be?

\section*{Acknowledgments} 
We thank Nathan Benjamin, Nathan Berkovits, Scott Collier, Sergei Dubovsky, Tom Hartman, Davide Gaiotto, Juan Maldacena, Dalimil Mazac, Joao Penedones, Eric Perlmutter and David Simon-Duffins, for useful 
comments and discussions.
JA also thanks Suzanne Bintanja, Jeevan Chandra and Gabriele Di Ubaldo for discussions.
FA also thanks Kostas Skenderis, Marika Taylor and David Turton for discussions.
Research at Perimeter Institute is supported in part by 
the Government of Canada through the Department of Innovation, Science, and Economic
Development Canada and by the Province of Ontario through the Ministry of Colleges
and Universities. 
RCM  and PV are supported in part by Discovery Grants from 
the Natural Sciences and Engineering Research Council of Canada, 
and by the Simons Foundation through the ``It from Qubit'' 
and the ``Nonperturbative Bootstrap'' collaborations, respectively  (PV: \#488661).
This work was additionally supported by   
FAPESP Foundation through the grants 2016/01343-7, 2017/03303-1, 2020/16337-8.
JA, FA and PV thank the organizers of the 
conference ``Gravity from Algebra: Modern Field Theory 
Methods for Holography", and acknowledge  KITP for hospitality. 
Research at KITP is supported 
in part by the National Science Foundation under Grant No. NSF PHY-1748958.
FA is supported by the Ramon y Cajal program through the fellowship RYC2021-031627-I funded 
by MCIN/AEI/10.13039/501100011033 and by the European Union NextGenerationEU/PRTR.

\appendix

\section{Solutions of AdS$_3$ Einstein Gravity}
\label{RobertsAppendix}

A striking consequence of the simplicity of gravity 
in three dimensions is that the general asymptotically 
AdS$_3$ solution can be written in closed  Ba\~nados form given in eq.~\reef{eq:Banados}. That is, 
\beq
\label{eq:BanadosA}
ds^2 = \frac{dy^2+dz d\bar{z}}{y^2} + L(z) dz^2 + 
\bar{L}(\bar{z}) d\bar{z}^2 + y^2 L(z) \bar{L}(\bar{z}) dz d\bar{z}\,.
\eeq
This solution is determined in terms of a holomorphic 
function $L(z)$ and an anti-holomorphic function $\bar{L}(\bar{z})$, 
which are proportional to the holomorphic and anti-holomorphic 
components of the boundary stress tensor.  The problem of 
finding solutions corresponding to configurations in the 
boundary CFT$_2$ therefore reduces to finding the boundary stress tensor \reef{barn7}.
Such an expansion is available in higher-dimensional Einstein gravity as well, but uniquely in three dimensions, the expansion truncates for the general solution.

\subsection{The det($g$) Wall}

While it is possible to construct a solution corresponding to any boundary stress tensor as an expansion near the conformal boundary, this expression will typically only describe the solution in a coordinate patch.  To see this, consider the determinant of the metric, which yields
\beq
\det(g) = \frac{\left(1-y^4 L(z) \bar{L}(\bar{z})\right)^2}{4 \ y^6}
\eeq
So we see that the metric degenerates when the numerator vanishes, which happens at
\beq
y = \left( L(z) \bar{L}(\bar{z}) \right)^{-1/4}
\eeq
Note that when the boundary stress tensor has poles, as it does when there are local operator insertions in the CFT$_2$, this surface will extend all the way to the conformal boundary at $y=0$.  In order to calculate the bulk action, we first need to extend the geometry past this surface.

\subsection{Extending the Solution}

Since every solution to Einstein gravity is locally isometric, there exists a change of variables that puts the metric of the general solution (\ref{eq:BanadosA}) into the form of the standard Poincar\'e upper half plane metric on Euclidean AdS$_3$,
\beq
ds^2 = \frac{dY^2 + dZ d\bar{Z}}{Y^2}
\eeq

The task of extending the solution can be informed by considering the explicit form of this map from the general solution to Euclidean AdS$_3$.  Remarkably, the explicit expression for this map is known \cite{Roberts:2012aq}.
The map is given explicitly by
\beq
\label{RobertsInAppendix}
\begin{split}
Y &= y\, \frac{4 (\partial f\,\bar{\partial} \bar{f})^{3/2}}{4\, \partial f\,\bar{\partial}\bar{f} + y^2 \,\partial^2 f\, \bar{\partial}^2 \bar{f}}\,, \\
Z &= f(z) - \frac{2 y^2 \,(\partial f)^2 \,\bar{\partial}^2\bar{f}}{4\, \partial f\,\bar{\partial}\bar{f} + y^2 \,\partial^2 f\, \bar{\partial}^2 \bar{f}}\,,  \\
\bar{Z}  &= \bar{f}(\bar{z}) - \frac{2 y^2\, (\bar{\partial}\bar{f})^2\, \partial^2 f}{4\, \partial f\,\bar{\partial}\bar{f} + y^2 \,\partial^2 f\, \bar{\partial}^2 \bar{f}}\,.
\end{split}
\eeq
Here the functions $f(z)$ and $\bar{f}(\bar{z})$ are related to the boundary stress tensor through the Schwarzian equations
\beq
\begin{split}
&\{f,z\} \equiv \frac{f'''(z)}{f'(z)} - \frac{3}{2}\left(\frac{f''(z)}{f'(z)}\right)^2 = -2 L(z)\,
\\
&\{\bar{f},\bar{z}\} = -2 \bar{L}(\bar{z})\,.
\end{split}
\label{eq:Schwarzian}
\eeq
At the conformal boundary $y=0$, we have
\beq
Z = f(z) \quad; \qquad
\bar{Z} = \bar{f}(\bar{z})
\,.
\eeq

The functions $f$ and $\bar{f}$ will generically have branch points, and the image under the map (\ref{RobertsInAppendix}) will be a region in Euclidean AdS$_3$ bounded by surfaces that are the images of surfaces on either side of the branch cuts.
These surfaces are identified with each other to produce the solution (\ref{eq:BanadosA}).
To find the completion of this solution, we need to extend the identification surfaces into the rest of AdS$_3$.
The map into AdS$_3$ is not necessarily injective.
So the completion of the solution (\ref{eq:BanadosA}) can be thought of as a subregion of a covering space of Euclidean AdS$_3$ with its boundary glued together in a way that can be ``read off'' from the functions $f$ and $\bar{f}$.

\begin{figure}
    \centering
    \includegraphics[scale=0.4]{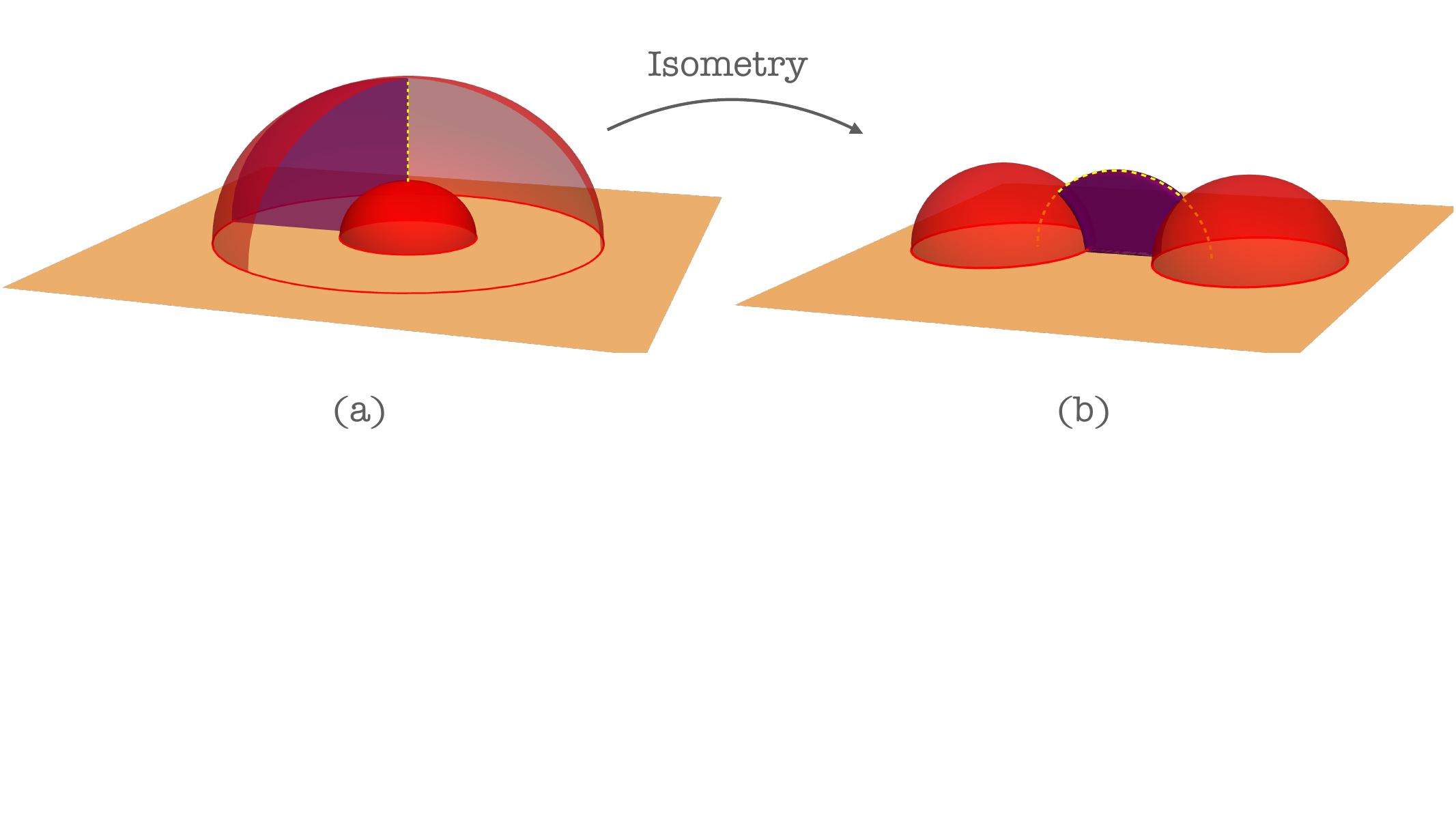}
    \vspace{-4cm}
    \caption{Replacing $f\rightarrow \frac{a f+b}{c f+d}$ acts on the domes-and-doors pictures as an isometry of Euclidean AdS$_3$. For example, here we show the effect of such a change on the representation of the two-point function geometry. Since we have only changed things by an isometry, the resulting geometry is the same.}
    \label{isometryInApp}
\end{figure}

Note that the functions $f$ and $\bar{f}$ are only unique up to isometries of Euclidean AdS$_3$.
Given solutions, $f_0$ and $\bar{f}_0$, to the Schwarzian equations (\ref{eq:Schwarzian}), we can construct a family of solutions
\beq
\begin{split}
f(z) = \frac{a f_0(z) + b}{c f_0(z) + d} \\
\bar{f}(\bar{z}) = \frac{\bar{a} \bar{f}_0(\bar{z}) + \bar{b}}{\bar{c} \bar{f}_0(\bar{z}) + \bar{d}}
\end{split}
\eeq
When we perform such a M\"obius transformation on $f$, the Roberts map \eqref{RobertsInAppendix} extends this change into the bulk.

The Roberts map with the new $f$ is the same as the old Roberts map, composed with an isometry of Euclidean AdS$_3$.
For example we can perform such an isometry by replacing $f_{\texttt{2pt}}$ with a new $f$ related by a M\"obius transformation in order to change the domes-and-doors picture from figure \ref{isometryInApp}(a) to figure \ref{isometryInApp}(b).
The freedom to perform such M\"obius transformations can be useful for putting the domes-and-doors constructions into nice standard configurations, such as the one in figure \ref{3ptIdents2dOnline}.

\section{Black Hole Two-Point Function in AdS$_3$}\label{2ptApp}

The black hole two-point function was treated in \cite{Paper1}.  Here, we perform the same calculation, in AdS$_3$, in a language similar to that which we used to treat the three-point function.  This allows us to determine the correct counterterms to add in the case of the three-point function.

\subsection{Geometric Setup}

We consider the black hole two-point function geometry, which can be described using the metric
\beq
ds^2 = d\tilde{\rho}^2 + \cosh^2\!\tilde{\rho}\,d\tau^2 + \sinh^2\!\tilde{\rho}\, d\theta^2
\eeq
where the coordinates range over
\beq
\begin{split}
0 < \tilde{\rho} <\infty\,, \\
0 < \tau < 2\pi R_h\,, \\
-\infty < \theta < \infty\,.
\end{split}
\eeq

We will calculate the Einstein-Hilbert action of this geometry by cutting off the radial coordinate at
\beq
\tilde{\rho}_{\text{cutoff}} = \ln \left(\frac{2}{\epsilon}\right) - \frac{\tilde{\varphi}(z,\bar{z})}{2}\,,
\eeq
as described in section \ref{Action}.
As in eq.~\reef{ace8}, the coordinates $z,\bar{z}$ are related to the coordinates $\tau,\theta$ with
\beq
\tau = \frac{1}{2}\left(\tilde{f}(z) + \bar{\tilde{f}}(\bar{z})\right)\quad;
\qquad
\theta = \frac{1}{2 i}\left(\tilde{f}(z) - \bar{\tilde{f}}(\bar{z})\right)\,.
\eeq
Further, the solution to the Laplace equation is given by
\beq
\tilde{\varphi}(z,\bar{z}) = \ln\left(\partial\tilde{f}\,\bar{\partial} \bar{\tilde{f}}\right)\,.
\eeq
With these choices, we have that the metric on the cutoff surface is
\beq
ds^2_{cutoff} = \frac{dz d\bar{z}}{\epsilon^2} + O(\epsilon^0)
\eeq

\subsection{Calculating the Action}

The contributions to the action coming from the bulk and the asymptotic boundary for our solution are given by
\beq
I_0 = \frac{1}{4 \pi \GN}\int_\mathcal{M} d\tilde{\rho} \ d^2z \sqrt{g} - \frac{1}{8 \pi \GN} \int_{\partial\mathcal{M}} d^2z \sqrt{h} = \frac{1}{4 \pi \GN} \left(V - \frac{1}{2} A\right)
\eeq
Now, we have
\beqa
\frac{1}{4 \pi \GN} \int_0^{\tilde{\rho}_{cutoff}} d\tilde{\rho} \sqrt{g} &=& \frac{1}{8 \pi \GN\,\epsilon^2} - \frac{e^{\tilde{\varphi}(z,\bar{z})}}{16 \pi\GN} + O(\epsilon^2)\,,
\\
-\frac{1}{8 \pi \GN} \sqrt{h} &=& -\frac{1}{8 \pi \GN\, \epsilon^2} - \frac{\partial \tilde{\varphi} \bar{\partial} \tilde{\varphi}}{16 \pi \GN} + O(\epsilon^2)\,.
\nonumber
\eeqa
Combining these results, we find
\beq
I_0 = \frac{1}{16 \pi \GN} \int d^2 z \left(-\partial \tilde{\varphi} \bar{\partial}\tilde{\varphi} - e^{\tilde{\varphi}(z,\bar{z})}\right)\,.
\eeq
As discussed in \cite{Paper1}, we must also add a Gibbons-Hawking-York boundary term on the stretched horizon $\tilde{\rho} = \epsilon$, which yields
\beq
I_{\text{GHY,hor}} = \frac{1}{8 \pi \GN} \int d^2 z \ e^{\tilde{\varphi}(z,\bar{z})}\,.
\eeq
Finally, we get
\beq
I = I_0 + I_{\text{GH,hor}} = \frac{1}{16 \pi \GN} \int_{\partial \mathcal{M}} d^2z \left(-\partial \tilde{\varphi} \bar{\partial} \tilde{\varphi} + e^{\tilde{\varphi}(z,\bar{z})}\right)
\eeq
Note that using the equations of motion, we have $\partial \tilde{\varphi} \bar{\partial}\tilde{\varphi}=\bar{\partial}( \tilde{\varphi} \partial \tilde{\varphi})$.
So we see that
\beq
\int d^2 z \  \partial \tilde{\varphi} \bar{\partial}\tilde{\varphi} = \int d^2 z \ \bar{\partial}( \tilde{\varphi} \partial \tilde{\varphi}) = i\oint dz \ \tilde{\varphi} \partial \tilde{\varphi}
\eeq
and therefore
\beq
I = \frac{1}{16 \pi \GN} \int d^2 z \left(\partial \tilde{\varphi} \bar{\partial} \tilde{\varphi} + e^{\tilde{\varphi}(z,\bar{z})}\right) + \frac{i}{8 \pi G} \oint dz \ \tilde{\varphi} \partial \tilde{\varphi}\,.
\eeq

\subsection{Counterterms}

The Laplace solution corresponding to two hyperbolic singularities located at $z = \pm 1$ is given by
\beq
\tilde{\varphi}_{2pt}(z,\bar{z}) = \ln\left(\frac{4 R_h^2}{|z^2 - 1|^2}\right)
\eeq
Inserting this into the action above, we find
\beq
I = \frac{1}{4 \pi \GN} \int d^2 z \left(-\frac{|z|^2}{|z^2 - 1|^2} + \frac{R_h^2}{|z^2 - 1|^2}\right)
\eeq
where we cut out a region of size $\epsilon$ around each of the insertions.  The $R_h$ dependent term gives
\beq
\frac{1}{4\pi \GN}\int d^2 z \frac{R_h^2}{|z^2-1|^2} = -\frac{1}{4 \GN} R_h^2 \ln(\epsilon)\,.
\eeq
So we see that
\beq
I = -\frac{1}{4 \GN} R_h^2 \ln(\epsilon) + C
\label{act99}
\eeq
where the constant $C$ does not depend on $R_h$.

To fix $C$, we note that when $R_h^2 = -1$, we should have the vacuum.
The vacuum partition function can be calculated from the conformal anomaly \cite{Chandra:2022bqq}, and the result is
\beq
-\ln(Z(S^2)) = -\frac{1}{2 \GN} \ln\left(\frac{R}{\epsilon}\right)\,.
\eeq
So we have
\beq
I|_{R_h\rightarrow -1} = \frac{1}{4 \GN} \ln\epsilon + C = -\frac{1}{2 \GN} \ln\left(\frac{R}{\epsilon}\right)\,,
\eeq
which allows us to read off the value of the constant $C$,
\beq
C = -\frac{1}{2 \GN}\ln R + \frac{1}{4 \GN} \ln\epsilon\,.
\eeq
Finally, we find the action \reef{act99} becomes
\beq
I = \frac{1}{4 \GN}(1 - R_h^2)\ln\epsilon -\frac{1}{2 \GN} \ln R\,.
\eeq

Hence, to leading order, the action is given by
\beq
I \simeq \frac{1}{4 \GN}(1-R_h^2)\ln\epsilon\,,
\eeq
and so for each operator, we must add a counterterm of the form
\beq
I_{ct}(R_h) = -\frac{1}{8 \GN}(1-R_h^2) \ln\epsilon\,.
\eeq

\section{Negative Mass}\label{negativeM}

It is interesting to consider analytically continuing the two- and three-point geometries to include insertions with negative masses.
Consider first the two-point case, with operators of mass $M$ placed at $0$ and $\infty$ for convenience.
Recall that when the mass is positive, we encountered a ``wall'' at 
\beq
y = \frac{1}{(L \bar{L})^{1/4}} = 2 \sqrt{\frac{z \bar{z}}{|M|}}
\eeq
where $\det(g) = 0$.
Some of the geometry was hidden behind this wall, and we needed to extend beyond this point to evaluated the two-point function.
In the case where the mass is negative, we still have that the determinant of the metric vanishes at $y = 2 \sqrt{z \bar{z}/|M|}$.
However, when $M<0$, the coordinate patch below this value of $y$ actually represents the complete geometry.
To see this, one need only use the Roberts map (\ref{Roberts}) to check where the $\det(g) = 0$ wall is mapped to in the ``domes-and-doors'' picture.
We find that the wall is mapped into the line $Z = \bar{Z} = 0$.
So in the negative mass case, there is no region hidden behind the wall.
In fact, the wall is the location of the conical excess sourced by the operator insertions!

\begin{figure}
    \centering
    \includegraphics[scale=0.4]{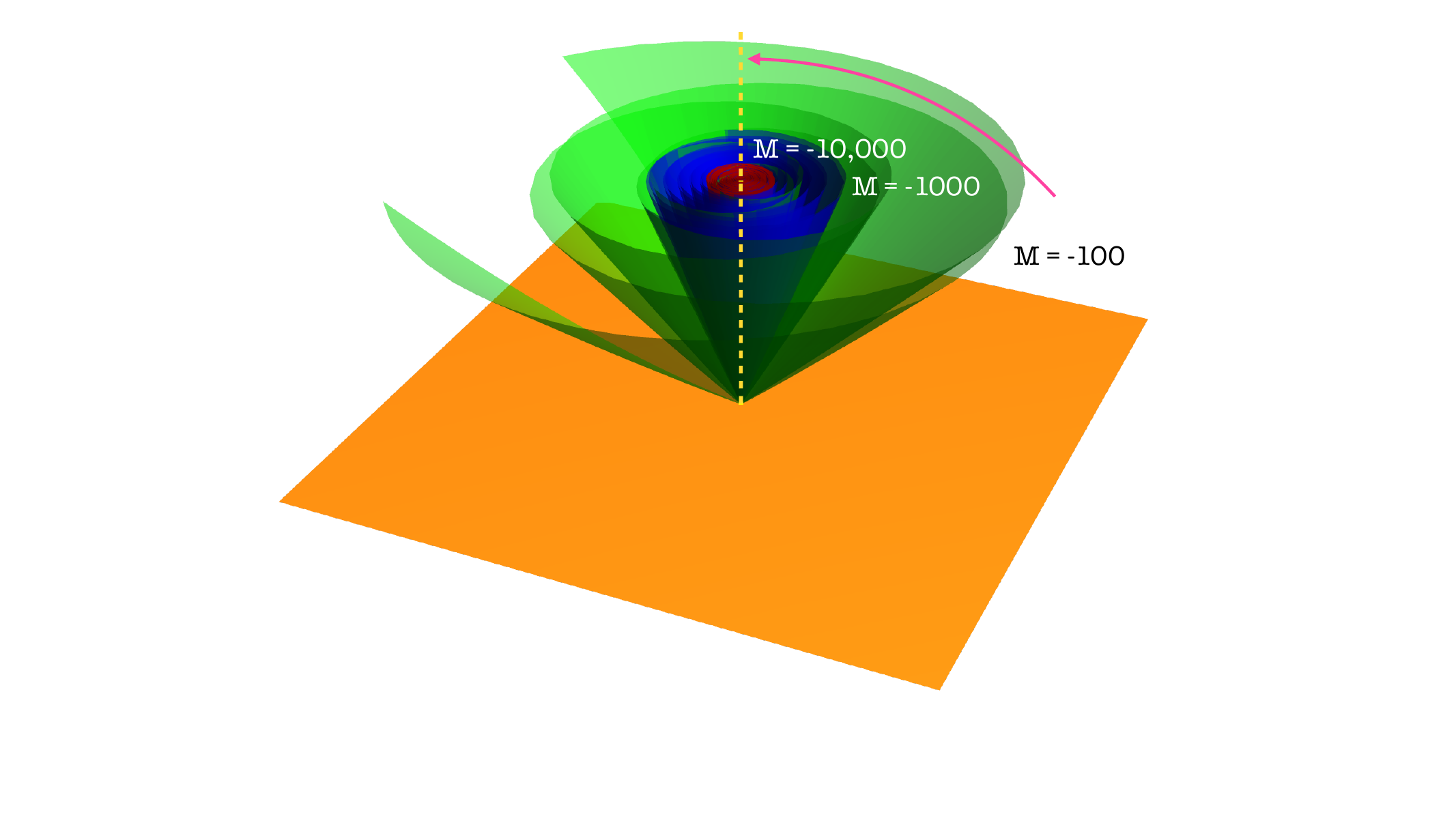}
    \vspace{-1.5cm}
    \caption{Here we show the image of the $\det(g)=0$ surface in empty AdS$_3$. We show the surface for $M=-100$ (green), $-1000$ (blue), and $-10,000$ (red). As the figure suggests, the wall shrinks towards the conical excess at $\tilde{\rho} = 0$ (the dashed yellow line in the figure) as we increase $|M|$. This shrinking is enough to compensate for the fact that the range of the angular coordinate $\theta$ is growing as $O(\sqrt{|M|})$, resulting in a volume that is only $O(M^0)$.}
    \label{negMFig}
\end{figure}

In order to calculate the two-point function with negative mass, we would like to consider the Ba\~nados metric \eqref{eq:Banados} with the operators at finite points.
In this case, the situation is slightly more complicated.
The Ba\~nados patch no longer covers everything.
However, if we consider the surface where $\det(g) = 0$, and use the map \eqref{Roberts} to identify it with a surface in empty AdS$_3$, we find that the resulting surface closes up towards the singularity as we increase the mass -- see figure \ref{negMFig}.
Indeed, we find
\beq
\tilde{\rho}_{\texttt{wall}}(\theta) = O\left(\frac{1}{(-M)^{1/4}}\right)
\eeq
where $\tilde{\rho}$ and $\theta$ are the coordinates introduced in eq.~\eqref{arc14}.
The range of $\theta$ in these coordinates is $0<\theta<2 \pi \sqrt{1-M}$.
So for large negative $M$, it is growing like $\sqrt{-M}$.
Hence we find that the volume of the region behind the $\det(g)=0$ wall is, to leading order at large negative $M$,
\beq
V_{\text{hidden}} = \int d\tilde{\rho} \, d\tau\, d\theta \cosh{\tilde{\rho}} \sinh{\tilde{\rho}} \sim \int  \tilde{\rho} \, d\tilde{\rho} \, d\tau \, d\theta \sim \pi \sqrt{1-M} \tilde{\rho}^2_{\texttt{wall}} \int d\tau = O(M^{0})
\eeq

On the other hand, consider the contribution from outside the $\det(g)=0$ wall in a solution described by the metric \eqref{eq:Banados} for arbitrary $L$, $\bar{L}$.
It is given by
\beq
I_{\texttt{outside}} = \frac{c}{6 \pi} \left(V_{\texttt{outside}}-\frac{1}{2}A\right)
\eeq
where $V_{\texttt{outside}}$ is the bulk volume outside the $\det(g) = 0$ surface, 
\beq
V_{\texttt{outside}} = \int_{y=\epsilon}^{y=1/(L\bar{L})^{1/4}} dy \, d^2 z \frac{1 - y^4 L(z)\bar{L}(\bar{z})}{y^3} = \int d^2 z \left(\frac{1}{2 \epsilon^2} - \sqrt{L \bar{L}} \right) + O(\epsilon^2)
\eeq 
and $A$ is the area of the cutoff surface at $y=\epsilon$,
\beq
A = \int d^2 z \frac{1}{\epsilon^2} + O(\epsilon^2)
\eeq
So we see that the contribution to the action from the patch covered by the expansion near the asymptotic boundary is given by
\beq\label{eq:sqrtLLb}
I_{\texttt{outside}} = -\frac{c}{6 \pi}\int d^2z \, \sqrt{L(z) \bar{L}(\bar{z})}\,.
\eeq
Now, returning to the specific case of the two-point function, we see that the action outside the $\det(g)=0$ wall is order $O(M^1)$.
So at large negative $M$, the leading contribution comes from outside the wall, and is given by simply \eqref{eq:sqrtLLb}.
We expect that a similar statement is true for the three-point function, since in that case, we find that the contribution outside the wall matches with the expected universal result at leading order in large negative mass.

\bibliographystyle{JHEP}
\bibliography{refs.bib}

\end{document}